\documentclass[letterpaper]{ar-1col}
\usepackage{rotating}%
\usepackage{etoolbox}
\usepackage[numbers,sort&compress]{natbib}%
\usepackage{amssymb,amsmath}
\usepackage{graphicx}
\usepackage{bm}

\newcommand {\subscript} [1] {\ensuremath{_{\textrm{#1}}}}
\newcommand {\SMB} {SmB\subscript{6}\ }

\newcommand {\Rxy} {R\subscript{xy}}
\newcommand {\Vxy} {V\subscript{xy}}

\newlength{\upit}\upit=0.1truein

\newcommand{\ltappr}{{{\lower4pt\hbox{$<$} } \atop \widetilde{ \ \ \ }}}
\newlength{\bxwidth}\bxwidth=1.5 truein

\newlength{\figwidth}
\newlength{\shift}
\usepackage{epsf}

\newcommand \bea {\begin{eqnarray} }
\newcommand \eea {\end{eqnarray}}

\newcommand{\beg}{\begin{equation}}
\newcommand{\en}{\end{equation}}

\newcommand{\dg}{^{\dagger }}
\newcommand{\bk}{\mathbf k}

\newcommand{\br}{\mathbf r}
\newcommand{\bG}{\mathbf G}

\jvol{6}
\jyear{2016}
\doi{10.1146/((please add article doi))}

\begin{document}

\markboth{Maxim Dzero et al.}{Topological Kondo Insulators}

\title{Topological Kondo Insulators}

\author{Maxim Dzero$^{1,2}$, Jing Xia,$^2$ Victor Galitski,$^{3,4}$ and Piers Coleman$^{5,6}$%
\affil{$^1$ Department of Physics, Kent State University, Kent, OH 44242, USA}
\affil{$^2$ Max Planck Institute for the Physics of Complex Systems, N\"{o}thnitzer str. 38, 01187 Dresden, Germany}
\affil{$^3$ Department of Physics and Astronomy, University of California, Irvine, California 92697, USA}
\affil{$^4$ Joint Quantum Institute, University of Maryland, College Park, Maryland 20742, USA}
\affil{$^5$ School of Physics, Monash University, Melbourne, Victoria 3800, Australia}
\affil{$^6$ Center for Materials Theory, Rutgers University, Piscataway, NJ 08854, USA}
\affil{$^7$ Department of Physics, Royal Holloway, University of London, Egham, Surrey TW20 0EX, UK}}

\firstpagenote{Accepted as an article in
the Annual Review of Condensed Matter Physics, Volume 7 (2016).}

\begin{abstract}
This article reviews recent theoretical and experimental work on a
new class of topological material - topological Kondo
insulators, which develop through the interplay of strong 
correlations and spin-orbit interactions. The history of Kondo
insulators is reviewed along with the theoretical models used to
describe these heavy fermion compounds. The Fu-Kane method of
topological classification of insulators is used to show that
hybridization between the conduction electrons and localized
f-electrons in these systems gives rise to interaction-induced
topological insulating behavior. Finally, some recent experimental
results are discussed, which appear to confirm the theoretical
prediction of the topological insulating behavior in Samarium
hexaboride, where the long-standing puzzle of the residual
low-temperature conductivity has been shown to originate from robust
surface states.
\end{abstract}


\maketitle

\tableofcontents

\section{INTRODUCTION}

{\sl Topological Kondo insulators} (TKIs) are a class of narrow gap 
insulator in which the gap is created 
by electron correlations, 
but which are at the same time, {\sl topologically ordered}. 
The first Kondo insulator (KI), \SMB was discovered
almost fifty
years ago~\cite{smb6} and today there are several 
known examples; at room temperature, these KIs
are metals containing a dense array of
magnetic moments, yet on cooling they 
develop a narrow gap
due 
the formation of {\sl Kondo singlets} which screen the 
local moments~\cite{revfiskaeppli,Fisk96,revki1,revki2}. 
{\sl Topological insulators} (TIs), discovered in just the last decade, 
are a new state of matter
~\cite{Hasan10,Qi11,kanemele05,bernevig06,Moore:2006uk,rahulroy09,kane1,konig07,hsieh08,kane2}:
in which a band-inversion gives bulk insulator with an 
energy gap that is traversed by a metallic Dirac surface state.

Kondo insulators were long 
regarded as a kind of ``renormalzed silicon'', with
a gap which is 
narrowed by the strong renormalizing effects of electron interactions~\cite{revfiskaeppli}.
The arrival of topological insulators forced a re-evaluation of this
viewpoint. The large spin orbit coupling, and the odd-parity of the
f-states led the current authors to propose
~\cite{Dzero2010} that Kondo insulators can become  topologically ordered.
The recent observation of robust~\cite{wolgasttki,kimtki} conducting
surface states in the oldest Kondo insulator \SMB  supports
one of the key elements of 
this  prediction, prompting a revival of interest 
in Kondo insulators as a new route for studying 
the interplay of strong interactions and topological order. 

In this article we review these recent developments, particularly
those surrounding SmB$_{6}$.  We begin by giving a brief review of the
early history and understanding of Kondo insulators, followed by a
review of recent developments associated
with these materials. 

\subsection{Key Properties of Kondo Insulators}\label{}

There are more than 
a dozen known Kondo insulators~\cite{revki2}, including the f-electron
materials \SMB ~\cite{smb6}, SmS under pressure~\cite{Maple:1971wq},
YbB$_{12}$~\cite{yb12},
CeFe$_{4}$P$_{12}$~\cite{cefep},
Ce$_{3}$Bi$_{4}$Pt$_{3}$~\cite{ce3bi4pt3}, CeRu$_{4}$Sn$_{6}$~\cite{ceru4sn6}
and the d-electron material
FeSi~\cite{jaccarino,fesi}. 
There are
also examples of ``failed'' Kondo insulators such as CeNiSn and
CeRhSb\cite{cenisn,cenisn-semimetal}
in which the insulating
gap appears to close in
certain directions\cite{ikeda,moreno} forming a semi-metal. 
Kondo insulators are the simplest example of 
heavy electron materials.  
At high temperatures, Kondo insulators are simply local moment metals, with
classic Curie-Weiss magnetic susceptibilities 
\begin{equation}\label{}
\chi (T)=
\frac{1}{3}\frac{\langle M^{2}\rangle }{T+\theta }
\end{equation}
that indicate the presence of a dense lattice of local moments.  
However, under the influence of the Kondo effect, 
the strength of the antiferrromagnetic interaction
between local moments and the conduction electrons grows, ultimately
leading to Kondo screening of the local moments to produce a paramagnetic
ground state.  In a simple picture, 
the narrow gap of Kondo insulators reflects the energy required
to break these emergent singlets. 

\subsection{Early History}\label{}

Until the 1970's magnetic materials containing f-electrons 
were thought to be electronically inert. 
The discovery of the first Kondo insulator 
SmB$_{6}$ by Anthony Menth, Ernest Buehler and Ted Geballe in 1969\cite{smb6} 
changed the perspective.  \SMB
is a paramagnetic metal at room temperature, with a
Curie-Weiss susceptibility characteristic of magnetic Sm$^{3+}$
ions, yet on cooling, it evolves into a paramgnetic
insulator with a tiny 10meV gap.  The discovery of similar behavior in
pressurized SmS led Brian Maple and Dieter
Wohlleben~\cite{Maple:1971wq,hirst}
to propose 
that coherent valence fluctuations in rare-earth ions
destabilize magnetism, allowing the f-electrons to delocalize into
the conduction sea. 

Building on these ideas, 
Neville Mott,  
Chandra Varma and Yako Yafet proposed the idea that 
Kondo insulators involve a kind of excitonic ordering between
localized f-electrons and delocalized d-
electrons~\cite{Mott:1974ui,varmayafet,Varma:1976wy}, giving rise to a hybridized 
band-structure with a gap. 
In another development, 
Sebastian Doniach introduced the 
the concept of a
``Kondo lattice''~\cite{doniach}: a lattice of localized moments
immersed in a sea of mobile electrons, described by the model 
\begin{equation}\label{klmodel}
H=-t\sum_{(i,j)\sigma } (c\dg_{i\sigma }c_{j\sigma }+{\rm H.c})
+ {J}\sum_{j, \alpha \beta } \vec{\sigma }_{j}
\cdot\vec S_{j}.
\end{equation}
Here $\vec{ \sigma }_{j}\equiv ( c \dg _{j\beta }\vec{\sigma }_{\beta \alpha }c _{j\alpha }
)$ is the spin density at site $j$ and $J$ is the antiferromagnetic 
Kondo coupling, and the spin $S=1/2 $ local moment, $\vec{S}_{j}$
at each site $j$.  
Doniach pointed out in the lattice, the physics is determined by a
competition between the Kondo
effect, which tends to screen the local moments, forming Kondo singlets
below 
the characteristic 
{\sl Kondo temperature}
\begin{equation}\label{}
T_{K}=D\sqrt{J\rho } \exp \left[ - \frac{1}{2J\rho }\right].
\end{equation}
and the  magnetic RKKY  (Rudderman Kittel Kasuya Yosida)
interaction between the local moments, 
which leads to magnetic order at the characteristic temperature $T_{RKKY} \sim J^{2}\rho .$
Doniach proposed that provided $T_{K}$ exceeds $T_{RKKY}$, the 
Kondo effect will overcome magnetism, to produce a singlet
ground-state.

\subsection{Strong Coupling model for the Kondo insulator}\label{}

\begin{figure}[h]
\includegraphics[width=3.4in,angle=0]{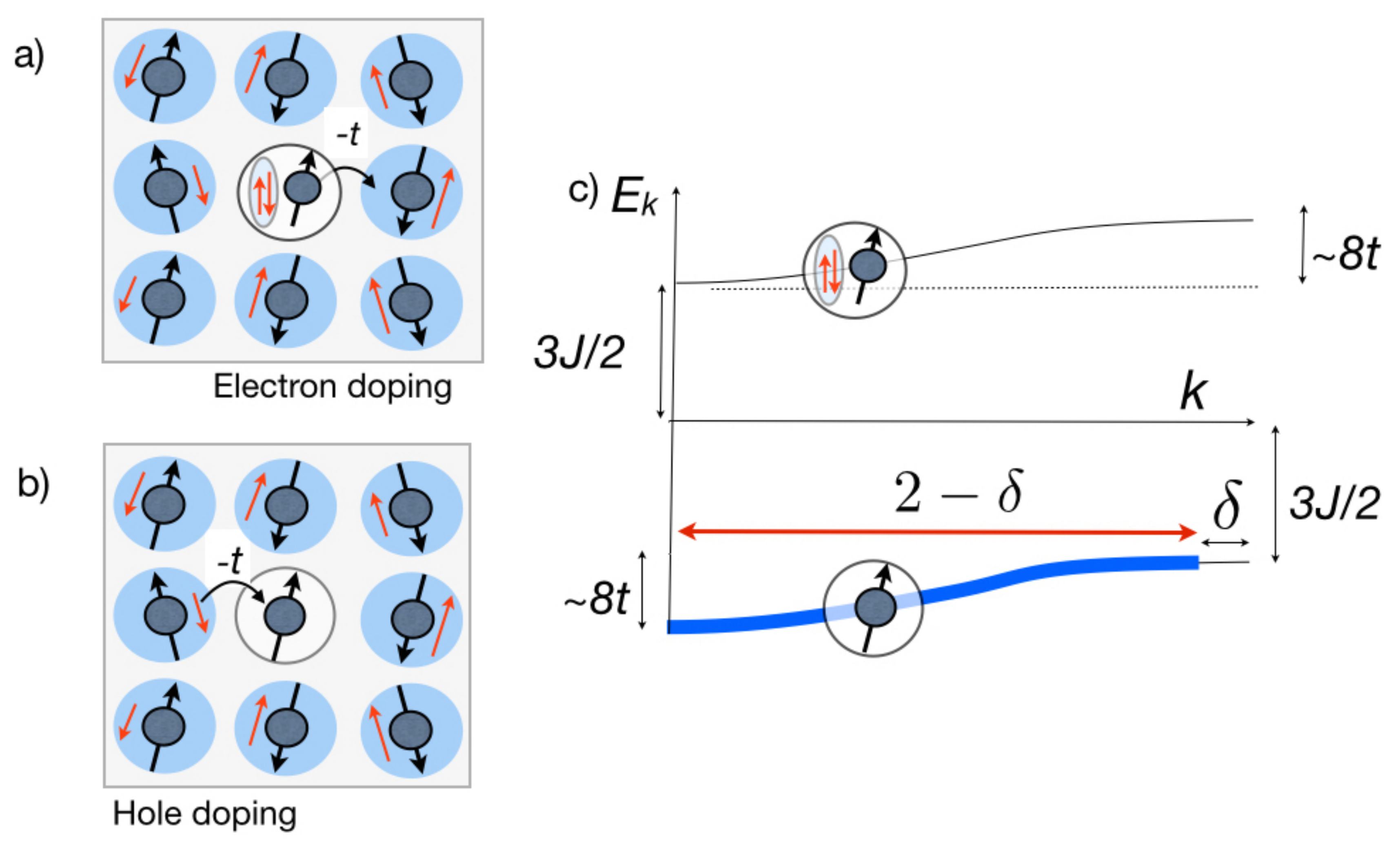}
\caption{Showing (a) electron and (b) hole
doping of strong coupling Kondo insulator. (c) Dispersion of strong
coupling Kondo insulator showing the formation of a heavy Fermi
surface when the Kondo insulator is hole-doped.}
\label{kondoinsfig2}
\end{figure}

A simple picture of the Kondo insulator is obtained by considering 
the Kondo lattice at strong coupling. 
Since 
the Kondo effect causes the coupling constant $J$ to renormalize to strong
coupling, the essence of the Kondo lattice
can be understood by examining 
the strong coupling limit in which $J$ is much larger than the hopping
$t$. 
In this limit,  the intersite hopping shown in 
(\ref{klmodel}) is a perturbation 
to the onsite Kondo insteraction, 
\begin{equation}\label{}
H\stackrel{t/J\rightarrow 0}\longrightarrow
{J}\sum_{j, \alpha \beta } 
\vec{\sigma }_{j} 
\cdot\vec S_{j}+O (t),\
\end{equation}
and the corresponding ground-state 
corresponds to the formation of a spin 
singlet at each site, denoted by the wavefunction
\begin{equation}\label{}
\vert KI \rangle=\prod_j \frac{1}{\sqrt{2}}\biggl( \Uparrow_{j} \downarrow_{j}-\Downarrow_{j}
\uparrow_{j} \biggr) 
\end{equation}
where the double and single arrows denote the localized moment and
conduction electron respectively, as illustrated in
Fig. \ref{kondoinsfig2} (a).

Each singlet
has a ground-state energy $E = - \frac{3}{2}J$ per site and a singlet-triplet
spin gap of magnitude $\Delta E = 2J$. If 
we remove an electron from site $i$, 
we break a Kondo singlet and 
create an unpaired spin with excited energy  $\frac{3}{2}J$, 
\begin{equation}\label{}
|\hbox{qp}^{+},i \uparrow \rangle  =  \Uparrow_{i}\prod_{j\neq i} \frac{1}{\sqrt{2}}\biggl( \Uparrow_{j} \downarrow_{j}-\Downarrow_{j}
\uparrow_{j} \biggr),
\end{equation}
while if we add an electron, the Kondo singlet is broken to 
create an electron quasiparticle,
a composite involving an unpaired local moment and a doubly occupied
conduction electron orbital
\begin{eqnarray}\label{l}
|\hbox{qp}^{-},i \uparrow \rangle  =
 \Uparrow_{i}
\biggl(\uparrow_{i}\downarrow_{i}\biggr)
\prod_{j\neq i} \frac{1}{\sqrt{2}}\biggl( \Uparrow_{j}
\downarrow_{j}-\Downarrow_{j}\uparrow_{j} \biggr), 
\end{eqnarray}
as illustrated in Fig \ref{kondoinsfig2}(b).
In this fashion, the strong coupling
Kondo lattice with one electron per site
forms an insulator with a charge gap of size $3J$ and a
spin gap of size $2J$.

If we subsequently reintroduce the hopping $-t$ between sites, then the
quasiparticles become mobile, as illustrated in
Fig. \ref{kondoinsfig2}.  
Thus if we hole- or electron- dope the Kondo
insulator by removing electrons, we end up with a narrow band of heavy
quasiparticles or {\sl heavy electrons}. In this way, Kondo insulators
can be regarded as the parent compounds of heavy fermion metals. 

\subsection{Adiabatic picture}\label{}

\begin{figure}[h]
\includegraphics[width=4.5in,angle=0]{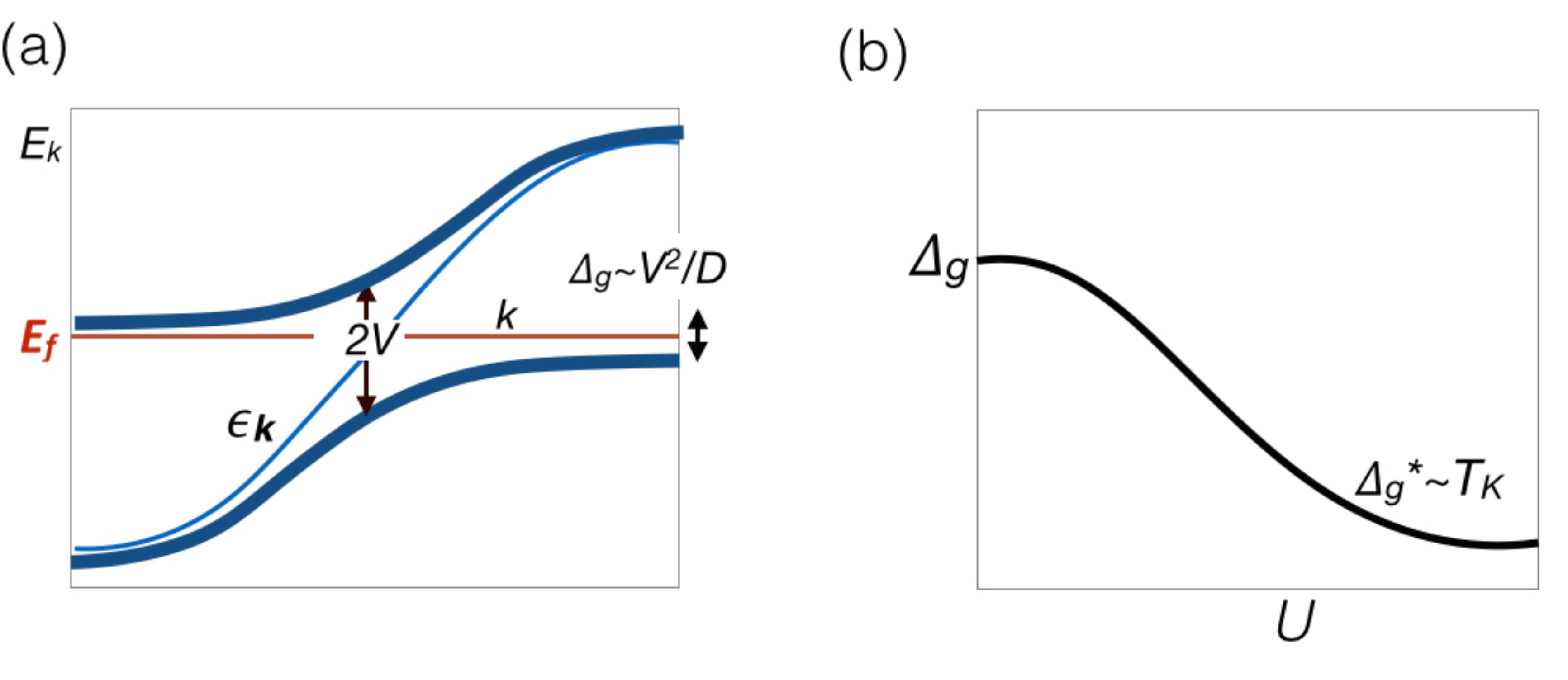}
\caption{Showing (a) hybridization of localized f and mobile d-band
gives rise
to a hybridization gap insulator with direct gap $2V$ and indirect gap
$\Delta_{g}\sim V^{2}/D$. (b) When the interaction is turned on
adiabatically, the band-gap renormalizes down towards the Kondo
temperature.}
\label{hybridization}
\end{figure}

An alternative way to understand Kondo insulators
in terms of {\sl adiabaticity} was proposed by 
Richard
Martin and Jim Allen~\cite{MartinAllen1}. 
The Kondo lattice Hamiltonian is a low energy limit
of the Anderson lattice model,
\begin{equation}\label{}
H_{ALM}= H_{c}+H_{f}+H_{hyb}
\end{equation}
where
\begin{equation}\label{}
H_{c}= -t\sum_{(i,j)\sigma } (c\dg_{i\sigma }c_{j\sigma }+{\rm H.c})
\end{equation}
describes the conduction electrons, 
\begin{eqnarray}\label{l}
H_{atom} (j)=\sum_{j} \left[ E_{f}n_{f} (j)+ Un_{f\uparrow} (j) n_{f\downarrow } (j) \right].
\end{eqnarray}
describes the atomic Hamiltonian of a localized f-state at site j
in energy level $E_{f}$ and corresponding onsite Coulomb repulsive interaction $U$,
while 
\begin{equation}\label{}
H_{hyb}
= 
\sum_{j}V\left[c\dg_{j\sigma }f_{j\sigma }+ {\rm H.c}
\right],
\end{equation}
describes the hybridization 
between the localized $f$ state
and the conduction electrons.   At $U=0$ and half filling, 
this model describes a simple hybridized
band-structure with a direct hybridization gap $V$ and an indirect gap
$\Delta_{g}\sim V^{2}/D$, where $D$ is the half band-width, 
as first noted by Mott~\cite{Mott:1974ui}. 
(See Fig. \ref{hybridization} (a))
By appealing to adiabaticity, Allen and Martin argued that 
as $U$ is increased, provided neutrality is maintained, the gap will 
simply renormalize downwards. 
At large  $U$,  onsite charge
fluctuations of the f-state can be eliminated via a canonical {\sl
Schrieffer Wolff} transformation~\cite{swolf} 
and in this limit, the model reduces
to the Kondo lattice model with $J\sim V^{2}/U$. 
Adiabaticity enables one to understand  the Kondo
insulator as simply the large $U$ cousin of the original hybridized
band insulator, with a duality between weak and strong coupling
in the two models:
\begin{eqnarray}\label{l}
\hbox{Large $U$ } &\leftrightharpoons& \hbox{Small $J\sim
\frac{V^{2}}{U}$ }\cr
\hbox{Small $U$}
& \leftrightharpoons& 
\hbox{Large $J\sim\frac{V^{2}}{U}$ }
\end{eqnarray}
The upshot of this discussion, is that
the low energy physics of the Kondo insulator can be equivalently 
described by a renormalized Anderson lattice model, with renormalized
parameters, $V^{*}$, $E_{f}^{*}$ and $\Delta_{g}^{*}$ determined by 
the Kondo temperature $T_{K}$ (See Fig. \ref{hybridization} (b)).
Today there are various methods for calculating these
renormalizations,  including 
path integral~\cite{lacroix, read83_1,auerbach}, slave boson
~\cite{barnes,slaveb,read83_v2,millislee,plasmon} 
Gutzwiller~\cite{brandow} and dynamical mean-field theory~\cite{krauth,coxdmft,jarrelldmft,hogandmft}
formulations of the Kondo and infinite $U$ Anderson models. 

\section{RISE OF TOPOLOGY}\label{}
The concept of topological order 
has its roots in the 
pioneering work of Robert Laughlin, of David Thouless, Mahito Kohmoto,
Peter Nightingale, Marcel den Nijs and of 
Duncan Haldane, on the integer quantum Hall effect or {\sl Quantum Hall
Insulator} \cite{Laughlin:1981jd,haldane88,tknn}. From this work, the
quantization of the Hall effect could be interpreted 
understood as a consequence of a {\sl topologically ordered}
ground-state wavefunction, in which the quantization of the Hall
constant results from the integer-valued Chern number of the topology.
Later work
of Shuichi Murakami, Naoto Nagaosa and Shoucheng Zhang, of Charles Kane
and Eugene M\' ele, and of
Andrei Bernevig and Taylor Hughes,
\cite{Murakami:2003tr,kanemele05,bernevig06} led to the concept of the
{\sl Spin
Hall insulator}: a two-dimensional topological insulator 
which is basically two time-reversed copies of the
Quantum Hall Insulator.  

In a series of remarkable papers, beginning around 2006,
Joel Moore and Leon Balents,~\cite{Moore:2006uk}, 
Liang Fu, 
Charles Kane and Eugene M\'el\'e,~\cite{kane1},
and Rahul Roy~\cite{rahulroy09}
made the discovery that three-dimensional
insulators can acquire a topological order through a spin-orbit driven
band-inversion, leading to a new kind of insulator. 
From this new perspective, the 
presence of topological order 
is determined by a single Ising or Z$_{2}$ index which is
positive in conventional insulators (Z$_{2}$=+1), but 
reverses sign in topological insulators (Z$_{2}$=-1). 
Conventional insulators can be loosely considered
as a {\sl miniature} version of the physical vacuum. By contrast,
topological insulators contain a internal twist to their
wavefunction that prevents them from being
adiabatically transformed into the vacuum of empty space,
and this is why they innevitably develop surface states. 
We can imagine smearing out the surface of an insulator so that the
the path from the insulator to
the vacuum of empty space becomes an adiabatic deformation of the
Hamiltonian. But if the vacua are topologically distinct, then the
gap can not remain open along this path, or it would be possible to
adiabatically deform the one into the other: as a result
the insulating gap must collapse at the interface between two
topologically distinct gapped states, to  produce a gapless surface
states. These states turn out to be Dirac surface states, with 
excitations whose spin is locked perpendicular to their momentum. 

Conventional surface
states are incredibly volatile and rarely survive as macroscopic
conducting surfaces, since they 
are highly sensitive to disorder, which leads to 
Anderson localization and surface reconstruction, which eliminates the
conducting states altogether. 
However, {\sl topologically protected } surface states
are robust against both Anderson localization and surface
reconstruction, leading to the unusual situation where even cracks
in the sample are conducting!
The  first two dimensional topological 
insulators were predicted in mercury cadmium-telluride (HgTe/CdTe)
quantum wells by 
Andrei Bernevig, Taylor Hughes and
Shou-Cheng Zhang~\cite{bernevig06} in 2006
and were 
discovered in 2007 by Lauren M\"olencamp and collaborators~\cite{konig07}. 
The first
experimentally realized three dimensional topological insulator was
bismuth antimonide (Bi$_{1-x}$Sb$_{x}$)~\cite{hsieh08}, 
discovered by the groups of Robert Cava and 
Zahid Hasan in 2008. 

In 2007,
Liang Fu and Charles Kane showed that if an insulator
has both time reversal and inversion symmetry 
~\cite{kane2}, 
the $Z_{2}$ index is uniquely 
determined by the 
the parities $\delta_{in}$ of the occupied Bloch states 
at the high symmetry points $\Gamma_{i}$ 
of the valence band 
\begin{equation}\label{}
Z_{2}= \prod_{\Gamma_{i}}\delta (\Gamma_{i}) = \left\{
\begin{array}{cl}
+1&\hbox{conventional insulator}\cr
-1&\hbox{topological insulator}
\end{array} \right.
\end{equation}
where $\delta (\Gamma_{i})=\prod_{n} \delta_{in}$ 
is the  the product of the parities of the
occupied states at the high-symmetry points
in the Brillouin zone. 
This beautiful index formula allows one to compute
whether an insulator  state is topological, merely by checking whether
$Z_{2}$ is negative.

\begin{figure}[h]
\includegraphics[width=4.5in,angle=0]{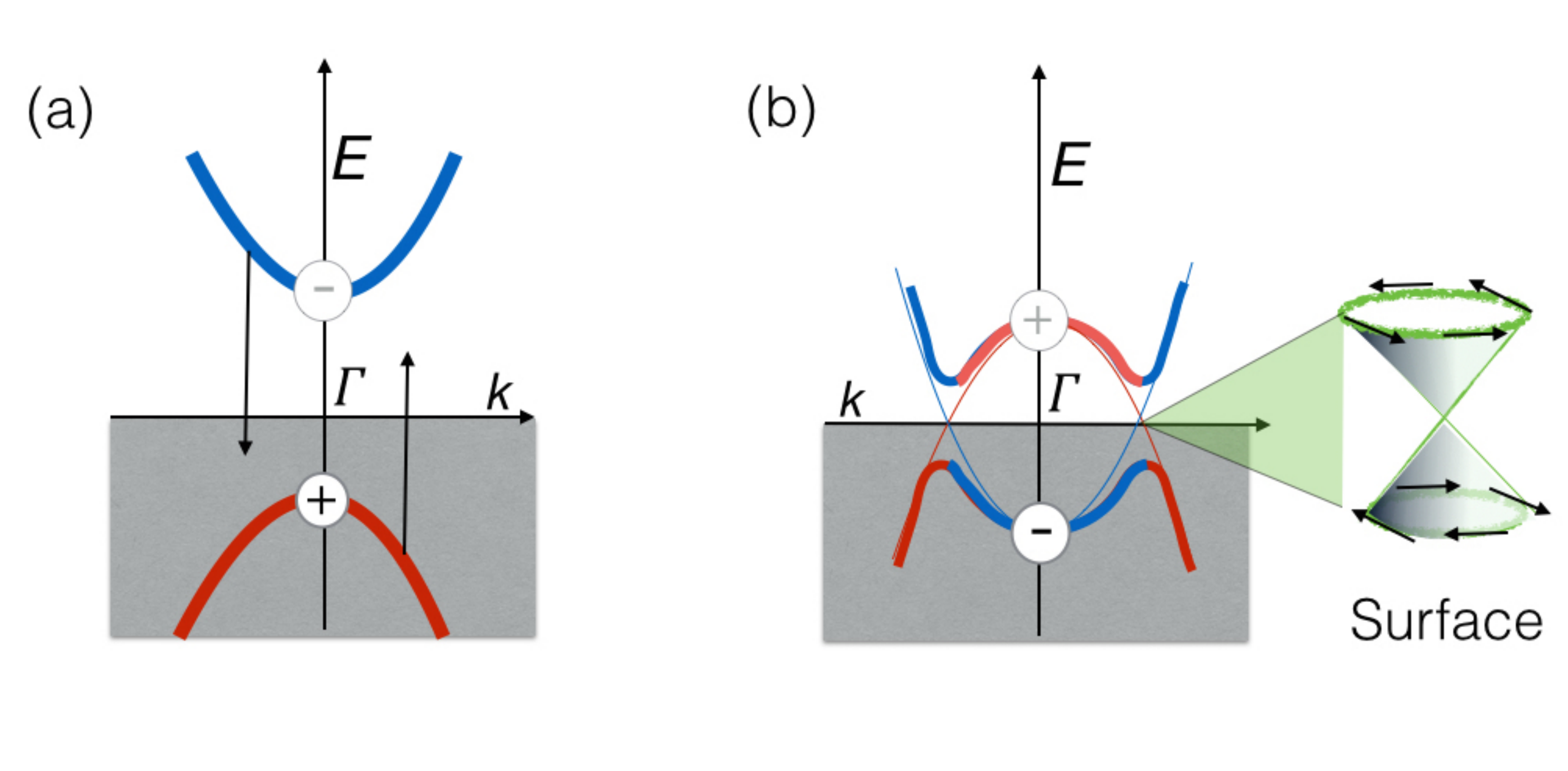}
\caption{Showing (a) topologically trivial band insulator with $Z_{2}=+1$ (b) 
band-crossing of even and odd parity states at an odd number of high
symmetry points 
leads to a topological insulator with $Z_{2}=-1$. Each band crossing
generates a Dirac cone of spin-momentum locked surface states. 
}
\label{band_crossing}
\end{figure}

Each time a band-crossing between an odd and even
parity state occurs, the $Z_{2}$ index changes sign (see
Fig. \ref{band_crossing}), and provided $Z_{2}=-1$, there will be
protected Dirac surface states.
For a cubic
insulator, with high symmetry points at the $\Gamma$,
$X$,  $M$ and $R$ points, this formula reduces to $Z_{2}= \delta_{\Gamma}
\delta_{X}\delta_{M}\delta_{R}$. (The X and M point occur three
times, but $\delta_{X,M}^{3}=\delta_{X,M}$).  For example, if $\delta_{X}=-1$
is the only negative parity, this means there have been {\sl three}
band crossings at the $X$ points, and in general this will give rise
to up to {\sl three} Dirac cone surface states.

\subsection{Topology meets strong correlation}\label{}
In 2010,  Maxim Dzero, Kai Sun, 
Victor Galitski
and Piers Coleman~\cite{Dzero2010} 
proposed that 
Kondo insulators  can 
form strongly interacting versions of the $Z_{2}$
topological insulator. 
The key points motivating this idea were that: 
\begin{itemize}

\item The spin orbit coupling of f-electrons 
in a Kondo insulator, of the order of $0.5$eV, is much larger than the
characteristic $10 meV$ gap of a Kondo insulator, making these 
essentially {\sl infinite spin orbit coupled} systems, ideal
candidates for spin-orbit driven topological order.

\item f-states are odd-parity, whereas the predominantly d-band 
conduction bands that hybridize with them 
are even parity, so that each time there is a
band-crossing between the two, the  $Z_{2}$ index changes sign, leading
to a topological insulator. 
\end{itemize}

The TKI proposal provides an appealing potential resolution of a
long-standing mystery in the Kondo insulator SmB$_{6}$, which for more
than thirty years, had been known to exhibit a low temperature
resistivity plateau ~\cite{allen79,cooley} (see Fig.~\ref{fig:ot}),
which could be naturally understood as a consequence of topologically
protected surface states~\cite{Dzero2010,Dzero2012}.  In 2012, teams
at the University of Michigan~\cite{wolgasttki} and the University of
California, Irvine~\cite{kimtki}, confirmed the existence of robust
surface states in SmB$_{6}$. Most recently 2014~\cite{Xu:1jv} Xu et
al. have detected the spin-polarized structure of the surface states
in these materials that tentatively confirm their topological
character (see discussion in section 4.2).

\section{THEORETICAL MODEL OF TOPOLOGICAL KONDO INSULATORS}

In this Section we review the models, which have been recently
proposed to describe various topologically nontrivial electronic
states in Kondo insulators. We begin by introducing 
a general
model for the topological Kondo insulators and discussing the
conjecture of adiabatic connectivity between band insulators and Kondo
insulators. This discussion is then followed by the review of the
tight-binding models specific to Samarium hexaboride.
\subsection{General discussion} The Anderson lattice model (ALM),
consisting of a set of localized $f$-electron states,
hybridized with conduction electrons, 
provides a basic description of the physics of $f$-electron
materials. 
The ALM Hamiltonian has
the form \beg\label{ALM}
\hat{H}_{ALM}=\hat{H}_f+\hat{H}_c+\hat{H}_{hyb}.  \en Here $\hat{H}_f$
describes the $f$-electron system and can be written as follows:
\beg\label{Hf}
\hat{H}_f=\sum\limits_{{ij}\alpha}t_{ij}^{(f)}\hat{f}_{i\alpha}\dg\hat{f}_{j\alpha}+\frac{U_{f}}{2}\sum\limits_{i\alpha\beta}
\hat{f}_{i\alpha}\dg\hat{f}_{i\alpha}\hat{f}_{j\beta}\dg\hat{f}_{j\beta},
\en where $t_{ij}^{(f)}$ are the hopping amplitudes for the
$f$-electrons between neighboring sites on the lattice,
$\hat{f}_{i\alpha}$ is a fermionic annihilation operator and
$U_{f}>0$ is the strength of the Coulomb repulsion between the
$f$-electrons on the same site.   The interactions between the
conduction electrons are ignored. 

In the simplest models of Kondo insulators
the $f$-electrons are 
considered as dispersionless, localized states,
 $t_{ij}^{(f)}=\epsilon_f \delta_{ij}$. However, in the
context of topological Kondo insulators, where the hybridization
contains nodes, a more general model
with nonzero off-diagonal elements of $t_{ij}^{(f)}$ is required 
(see our discussion below). Inside a crystal, the intrinsic degeneracy
of the $l=3$ $f$-states, 
$N=2 (2 l +1)= 14$ 
is lifted by the spin-orbit coupling and the crystalline fields,
and the subscript $\alpha$ in (\ref{ALM}) labels the components of the low-lying
spin-orbit split $f$-multiplet.

The second term in (\ref{ALM}) accounts for a band of non-interacting
conduction electrons: 
\beg\label{Hc}
\hat{H}_c=\sum\limits_{{ij}\sigma}\sum\limits_{a,b}t_{ia,jb}^{(c)}\hat{c}_{i,a\sigma}\dg\hat{c}_{j,b\sigma},
\en 
where $\sigma=\uparrow,\downarrow$ labels the projection of
conduction electron's spin, $(a,b)$ accounts for the degeneracy of the
corresponding conduction orbitals due to non-zero angular momentum and
$t_{ia,jb}^{(c)}$ is the hopping amplitude. Note that the intersite hopping
terms in (\ref{Hf}) and in (\ref{Hc}) are not 
necessarily limited to the nearest neighbor sites. Lastly, the
hybridization term in the Hamiltonian (\ref{ALM}) is 
\begin{eqnarray}\label{l}
\label{Hhyb}
\hat{H}_{hyb}=\sum\limits_{ij}\sum\limits_{a\sigma\alpha}\left(V_{ia\sigma,j\alpha}\hat{c}_{i,a\sigma}\dg\hat{f}_{j\alpha}+V_{i\alpha,ja\sigma}\hat{f}_{i\alpha}\dg\hat{c}_{j,a\sigma}\right)
\end{eqnarray}
with $V_{ia\sigma,j\alpha}$ being the hybridization matrix
diagonal in momentum space: \beg\label{Vk}
V_{ia\sigma,j\alpha}=\sum\limits_{\bk}V_{a\sigma,\alpha}(\bk)\exp[-i\bk\cdot(\br_i-\br_j)].
\en 
The important point, is that all the spin-orbit coupling, all the
topology
is hidden inside the hybridization matrix. 
The specific form of the hybridization matrix
$V_{a\sigma,\alpha}(\bk)$ will depend on the multiplet structure of
the f-electrons as well as on the
angular momentum of the conduction electrons. However, typically the
conduction electrons are composed of {\sl even parity} $l=2$ d-states,
hybridizing with the $l=3$ $f$-electrons, which have {\sl odd parity} 
~\cite{revki2}. It  follows that at  the high symmetry points in the Brillouin zone
$\bk = \bk_{\Gamma}$, parity is a good quantum number and as a result, the
odd-parity f-states and the even parity d- states can not mix, so that
$V (\bk_{\Gamma})=0$.  More generally, since the Hamiltonian must be
even under the parity operation, any hybridization matrix that mixes 
even and odd parity electron states must itself be 
an odd-parity function of momentum 
\beg\label{Vodd}
V_{a\sigma,\alpha}(-\bk)=-V_{a\sigma,\alpha}(\bk), \en i.e. the
elements of the hybridization matrix are \emph{odd} functions of
momentum. 
We can see this more detail by noting that at a
high symmetry point, the effect of reversing $\bk_{\Gamma}$ is the
same as shifting it by a reciprocal lattice vector $\bf G$, 
$-\bk_{\Gamma} =\bk_{\Gamma}+ {\bf G}$, and since the hybridization is
periodic in $\bf G$, i.e $V (\bk_{\Gamma}+ \bG)$, it follows that $V
(-\bk_{\Gamma} )= V (\bk_{\Gamma} + \bG)= V (\bk_{\Gamma} )$.  Comparing this with (\ref{Vodd} ),
it follows that $V (\bk_{\Gamma})=0$ as expected. 
The presence of nodes in the hybridization at the high symmetry points
$\bk_{\Gamma}$ is central to a description of topological Kondo
insulators. 

\begin{figure}[h]
\includegraphics[width=4.5in,angle=0]{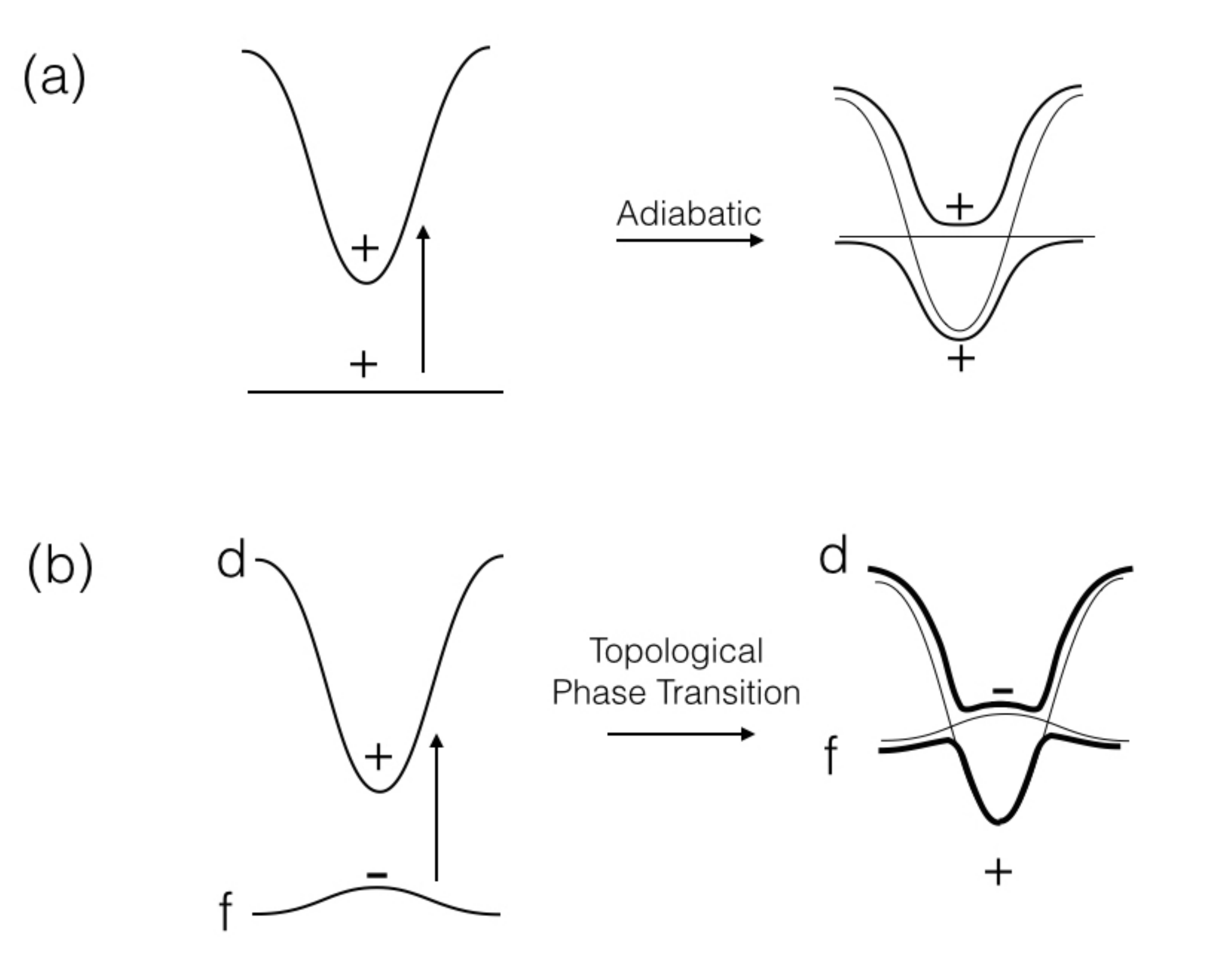}
\caption{(a) If we ignore the effects of topology 
in a conventional Kondo insulator, the interaction can be
turned on adiabatically. When the interactions are turned on, the
lower band is pushed into the upper band. 
Two bands of the same parity will always
repel one-another and will not cross when the interactions are turned
on. (b) When interactions are turned on in a topological insulator,
they can lead to band-crossing and a topological phase
transition. Here, interactions cause an f-band to push up into a
d-band. Since the two bands have opposite parity, they do not
hybridize at the high symmetry point so band-crossing occurs, leading
to a topological phase transition. 
}
\label{adiabatic}
\end{figure}

\subsubsection{Non-interacting limit and adiabatic continuity}
Let us consider the simplest case when the interaction between the $f$-electrons is zero, $U_{f}=0$. In this case, the model Hamiltonian
(\ref{ALM}) describes 
two hybridized bands of non-interacting electrons as illustrated in
Figure {\ref{hybridization}} (a), with a direct gap of order $V$ and
an indirect gap of the order of $|V|^2/D$ (where $|V|$ is the typical
size of the hybridization and $D$ is the conduction electron
bandwidth). 
The insulating ground state in this model
corresponds to the case of half-filling, so that the lower bands become fully occupied. For example, in the simplest case of a single-Kramers-degenerate conduction electron and $f$-orbital Kramers doublet, an insulating state is realized when there is exactly one conduction electron and one $f$-electron per site. 

Does the system remain insulating if we adiabatically switch on the
interaction $U_{f}$?  In their pioneering work Martin and Allen
\cite{MartinAllen1,MartinAllen2,Martin} argued using  Luttinger's
theorem, that as long as the band remained half-filled, there would 
be no reason for the gap to close (See Fig. \ref{adiabatic} a). 
Numerical
studies within the self-consistent fluctuation exchange approximation
have largely confirmed this conjecture at least for moderate values of
$U_{f}$.~\cite{revki2,Numerics} Theoretical schemes
based on the variational wave-functions  ~\cite{varmayafet}
also seem to suggest the validity of the Martin-Allen conjecture even for
$U_{f}\to\infty$.

However, topology
complicates adiabaticity  arguments~\cite{Dzero2010,Dzero2012}:
: in particular, as we saw in the last section, 
when the parities of the conduction and f-bands are opposite,
the hybridization vanishes high symmetry points, and this opens up the
possibility that interactions will induce
band-crossing, changing the topology of the ground-state. 
For example, let us assume that in the non-interacting
limit, the ALM Hamiltonian is topologically trivial, with a
completely filled band of f-states (See Fig. \ref{adiabatic} b). 
For a
system with time-reversal and inversion symmetry, the $Z_2$
topological invariant $\nu=0,1$ is determined by the parity operator
eigenvalues ~\cite{kane2}. For the ALM and taking into account
(\ref{Vodd}) it is simply given by 
\beg\label{Z2nu}
Z_{2}=(-1)^\nu=\prod_{m=1}^8\textrm{sign}[\epsilon_c(\bk_m)-\epsilon_f(\bk_m)],
\en 
where $\bk_m$ is a momentum at one of the eight high-symmetry
points of the 3D Brillouin zone and $\epsilon_{c,f}(\bk)$ is the
dispersion of the conduction and the $f$-electrons correspondingly.
As we switch the interaction $U_{f}$ adiabatically the conduction
d-band and f-bands will renormalize, with the f-level moving upwards
relative to the conduction bands due to their stronger Coulomb
interaction, 
$\epsilon_{c,f}(\bk)\to\tilde{\epsilon}_{c,f}(\bk)$. If as a result of
this process, the f- and d- bands cross at an odd number of points, 
then a topological phase transition will take place into a topological
insulator (see Fig. \ref{adiabatic} b). 
Since two topologically distinct states cannot be
adiabatically connected without closing the gap, this phase transition
must be accompanied by a momentary closing of the gap 
at a particular critical values of $U_{f}$. In this way, adiabaticity
can break-down. 
Recent studies by Werner and
Assaad ~\cite{DMFT} of the periodic ALM on the two-dimensional square
lattice have found that increasing the strength of the Hubbard
interaction leads to the series of transitions between normal
insulator (small $U_{f}$) into a strong topological insulator with the
band inversion at the $\Gamma$ point and then into another strong
topological insulator, in which bulk bands invert at the $X$ point of
the two-dimensional Brillouin zone.

From this discussion, it becomes clear that 
in order that the Allen-Martin idea of adiabaticity to
topological Kondo insulators, then the interaction and the strength of
the spin-orbit interaction in the Hamiltonian
must be tuned so that no band crossing takes place. 
Since interactions have the effect of narrowing 
the bands, in practice, this will mean starting with a Hamiltonian
with Hamiltonian with a greatly enhanced value of the spin-orbit
coupling.

\subsubsection{Model for topological Kondo insulators with Kramers doublets}
\begin{figure}[h]
\includegraphics[width=3.4in,angle=0]{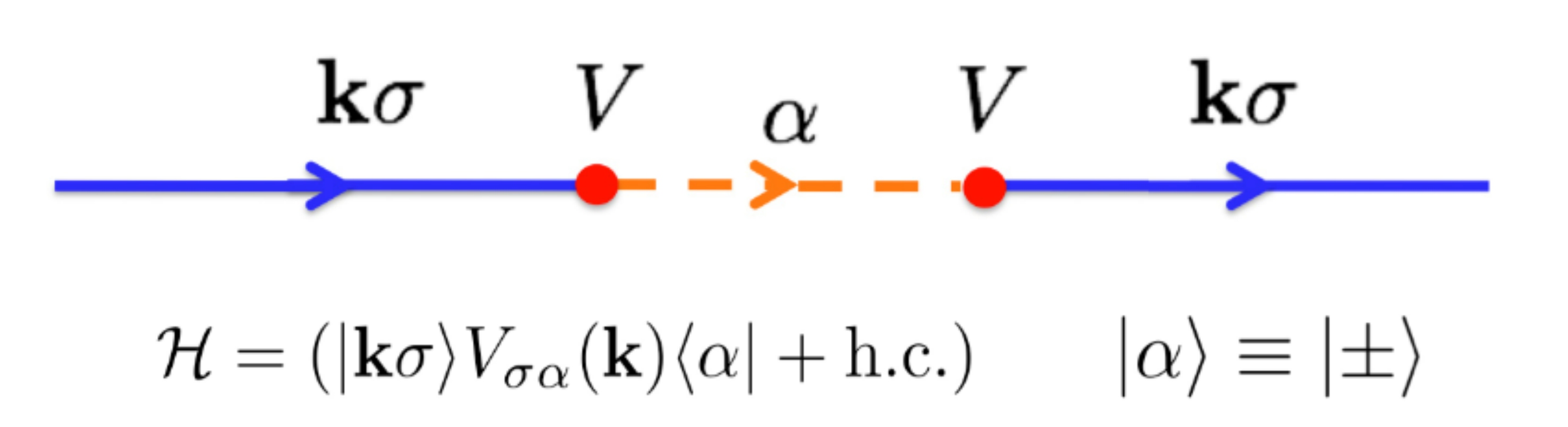}
\caption{Diagrammatic representation of the hybridization process between the conduction electrons and $f$-electron Kramers doublet.}
\label{FigV2}
\end{figure}

The simplest model for a 
topological Kondo insulator contains
a single $d$-band of conduction electrons hybridized
with a single Kramers f-doublet, with nearest neigbor
hopping and hybridization, Fig. \ref{FigV2}.  
The resulting non-interacting Hamiltonian has the form:
\beg\label{H0}
\hat{H}_0=\sum\sum\limits_{\bk ,\alpha\beta}\hat{\Psi}_{\bk\alpha}\dg 
\left(\begin{matrix} \epsilon_c(\bk) & 
V  {\vec{ d}} (\bk )\cdot
\vec{\sigma }
\\
V  {\vec{ d}} (\bk )\cdot
\vec{\sigma }
& 
\epsilon_f(\bk) \end{matrix}
\right)_{\alpha\beta}
\hat{\Psi}_{\bk\beta},
\en
where 
\beg\label{hk}
\epsilon_{c,f}(\bk)=-\frac{t_{c,f}}{6}(\cos k_x+\cos k_y+\cos
k_z)+\epsilon_{c,f}, 
\en
describe the dispersion of the conduction and f-band, while 
\begin{equation}\label{}
\vec{d} (\bk ) = (\sin k_{x},\sin k_{y},\sin k_{z})
\end{equation}
is a vector in
spin space, which approximates to $\vec{d} (\bk )\sim \bk$ in the
vicinty of the $\Gamma$ point. Notice how all the spin orbit coupling
effects are held by the hybridization.  
In the special case where $\epsilon_{c} (\bk )= - \epsilon_{f} (\bk
)$, this Hamiltonian is a simple lattice generalization 
of the mean-field Hamiltonian for the gapped topological superfluid phase of
Helium, He-3B\cite{vollhardt90,Balian:1963wq}.  
The non-interacting
band dispersion described by this Hamiltonian is simply 
\begin{equation}\label{}
E_{\bk \pm }= \frac{\epsilon_{c} (\bk)+\epsilon_{f} (\bk )}{2}
\pm 
\sqrt{
\left(
\frac{
\epsilon_{c} (\bk)-\epsilon_{f} (\bk )
}
{2}
\right)^{2}+ V^{2 }|\vec{d} (\bk )|^{2}
}.
\end{equation}
This model has been
employed to great success 
in a number of recent publications \cite{Sigrist,Erten15,Sigrist2,Vojta2}.

The low energy properties of the interacting model with the Hamiltonian
\beg\label{TKI}
\hat{H}=\hat{H}_0+\frac{U_{f}}{2}\sum\limits_{i\alpha\beta}
\hat{f}_{i\alpha}\dg\hat{f}_{i\alpha}\hat{f}_{j\beta}\dg\hat{f}_{j\beta}
\en
can be analyzed by either employing the conjecture that the effect of the local correlations between the $f$-electrons leads to the renormalization of the hybridization amplitude and the shift of the $f$-energy level (see e.g. ~\cite{Dzero2010,Dzero2012} and references
therein). Alternatively, one can set $U_{f}\to\infty$ and project out the doubly occupied $f$-states by introducing the slave-boson operators ~\cite{revki2,DzeroLargeN}. The slave-boson mean-field theory corresponds to replacing operators by $c$-numbers which are computed self-consistently. The result of the above mentioned procedure is a renormalization of the unperturbed bands (\ref{H0}):
\beg\label{lowEmodel}
V\to\tilde{V}=\sqrt{1-n_f}V, \quad t_f\to\tilde{t}_f=(1-n_f)t_f,
\en
where $n_f$ is an average $f$-level occupation number. 
Lastly, the position of the bare $f$-electron level $\epsilon_f$ shifts from below the chemical potential for the conduction electrons to 
energies $\epsilon_f+\lambda$ above it. Subsequent analysis of the topological invariant (\ref{Z2nu}) yields the odd number of band inversions for intermediate values of $n_f$ - mixed-valence regime - and even number of band inversions for $n_f\sim1$ - local moment regime.~\cite{Dzero2010,Dzero2012} The latter corresponds to the so called
``weak topological insulator'' with the topological invariant given by 
\beg\label{weaknu}
(-1)^{\nu_a}=\prod\limits_{\bk_m\in P_a}\textrm{sign}[\epsilon_c(\bk_m)-\epsilon_f(\bk_m)], \quad (a=x,y,z),
\en
where $P_a$ denotes a plane perpendicular to one of the main crystalline axes. 

An appearance of the weak topological insulating state for $n_f\sim 1$ raises a general question whether it would be in principle
possible to stabilize the strong topological insulating state ($\nu=1$) in the local moment regime. To address this issue Dzero 
~\cite{DzeroLargeN} has generalized conduction and $f$-electron Kramers doublets to $N$ components and used the symplectic SP($N$) ($N=2k, k=1,2,...$) representation for the electronic operators to properly account
for time-reversal symmetry of the conduction and $f$-electron states. In agreement with the previous results ~\cite{Dzero2010,Dzero2012,TKISU2} he found that for $N=2$ and $N=4$ there appears two (weak and strong) topologically non-trivial states depending on the relative position between the renormalized $f$-level and the chemical potential of the conduction band. Interestengly, for the large value of $N>4$ there is only strong topological insulating state. Thus, the higher degeneracy of the $f$-electron multiplet favors 
the strong topological state. This result has been later confirmed by considering the tight-binding model for Samarium hexaboride (SmB$_6$) with the fourfold degenerate $f$-orbital multiplet. ~\cite{Alexandrov13}

\subsection{Tight-binding models for Samarium hexaboride}
Samarium hexaboride has recently emerged as a prototypical candidate for the first experimental 
realization of strongly correlated topological insulator. 
Perhaps one of the most intriguing problems concerns the role the electronic correlations play in determining the 
parameters of the helical surface states such as the effective mass of the surface electrons and characteristic length scale on which 
surface states penetrate into the bulk. Thus, the formulation of the realistic and yet tractable tight-binding model presents
a first important step towards a better understanding of the physical properties of SmB$_6$. Below we review the recent theoretical models which have been put forward to describe the formation of the topological surface states in this material. 

\begin{figure}[h]
\includegraphics[width=3.4in,angle=0]{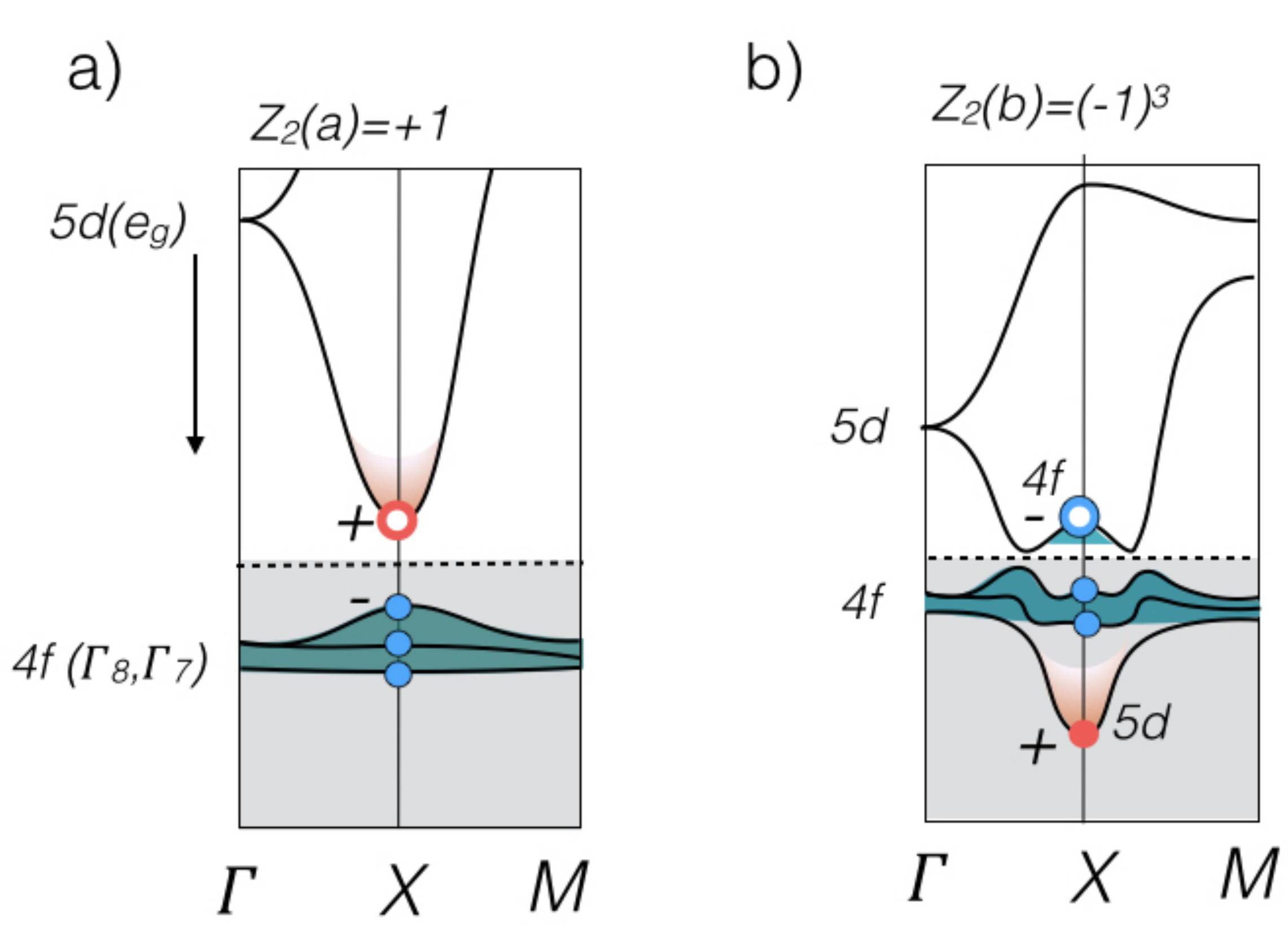}
\caption{Schematic illustration of the band-crossing between  d- and f- states
at the $X$ point in SmB$_{6}$. (a) Bands uncrossed.  The filled
4f$^{6}$ band of f-electrons is a conventional insulator. (b) Bands
crossed: the d-band cuts beneath the f-band at the $X$-point,
displacing an odd parity f-state from the valence band to the
conduction band. The resulting $(-1)^{3}$ sign reversal in the $Z_{2}$
index gives rise to a topological insulator.}
\label{FigSmX}
\end{figure}

The key insight for building 
realistic tight-binding model for SmB$_6$ came  from a detailed first-principle
calculations of Yanase and Horima ~\cite{Yanase} and of Antonov et
al. ~\cite{Antonov} Their calculations show that the 
Samarium $4f$ orbitals hybridize exclusively with 
Samarium $5d$-orbitals ~\cite{Yanase,Antonov}. 
The crystal field splitting of the 
$d$-multiplets leads to 
a low-lying $e_g$ doublet at the $\Gamma$ point. 
Away from the $\Gamma$ point, the 
$e_g$ orbitals split into two Kramers doublets, the lower one dipping
down at the $X$ point, where it dives through the 
the $4f$ bands. Hybridization between the two bands
forces 4f states from the valence to the conduction band, forming
heavy 4f electron band pockets at the X
points.  Once the
d-band crosses through the f-band at the three $X$ points, so long as
there are no other crossings, the resulting
non-interacting band-structure is innevitably topological,
{\sl independently of the details of the f-multiplets} (See Fig. \ref{FigSmX}).

Infact, in the cubic environment, the six $J=5/2$ 4f orbitals of the
Samarium split into a $\Gamma_{7}$ doublet and a $\Gamma_{8}$ quartet.
LDA studies ~\cite{Yanase,Antonov} suggest that the physics of the
$4f$ orbitals is governed by valence fluctuations involving electrons
of the $\Gamma_{8}$ quartet and the conduction $e_g$ states,
$e^{-}+4f^{5} (\Gamma_{8}^{(\alpha)})\rightleftharpoons4f^{6}$.  The
$\Gamma_8^{(\alpha)}$ ($\alpha=1,2$) quartet consists of the following
combination of orbitals:
$|\Gamma_8^{(1)}\rangle=\sqrt{\frac{5}{6}}\left\vert\pm\frac{5}{2}\right\rangle+
\sqrt{\frac{1}{6}}\left\vert\mp\frac{3}{2}\right\rangle,~
|\Gamma_8^{(2)}\rangle=\left\vert\pm\frac{1}{2}\right\rangle$. This
then leads to a simple physical picture in which the $\Gamma_{8}$
quartet of $f$-states hybridizes with an $e_g$ quartet of $d$-states
to form a Kondo insulator.

In 2011, Takimoto ~\cite{Takimoto2011} introduced a tight-binding model for
SmB$_6$ in which the hopping amplitudes in the Hamiltonian (\ref{ALM})
are non-zero for nearest- and next-nearest-neighbors, while
hybridization involves the nearest-neighbor overlap integrals only.
The values of the hopping amplitudes were adjusted to fit the 
LDA band structure results, while 
the effect of
interactions between the $f$-electrons is modelled as
a renormalization of the bare $f$-energy level and the hybridization. 
In Takimoto's model, 
a singlet $d$-like orbital inverts with an
$f$-like orbital at the X point of the bulk Brillouin
zone, while the remaining two bands remain inert.  This band inversion
at the X points implies the existence of three Dirac cones on the
surface: one at the surface $\overline{\Gamma}$ point and two at the
$\overline{X}$ points. Interestingly, the corresponding Fermi
velocities for the electrons at the $\overline{\Gamma}$ point are the
same, while the Fermi velocities at the $\overline{X}$ are strongly
anisotropic.~\cite{Takimoto2011}

Alexandrov and collaborators ~\cite{Alexandrov13} have considered a
simpler tight-binding model only nearest neighbor
hopping amplitudes only, showing that the LDA band structure results
are recovered by using the appropriate
Slater-Koster\cite{SlaterKosterfs} ratio
of the overlap
integrals between $d_{x^2-y^2}$ and $d_{3z^2-r^2}$ orbitals. By
setting $U_{f}\to\infty$ and employing the slave-boson mean-field
theory, it was shown that strong topological insulator state extends
all to way to the local moment regime, $n_f\sim 1$.  It was also
emphasized that quite generally, cubic topological insulator can only
be realized when bands invert at the $X$ or $M$ points, since at the
$\Gamma$ and $R$ points the $d$- and $f$-bands are doubly degenerate
and therefore remain topologically inert - the parity eigenvalues at
these points are always positive.

Legner and collaborators have considered an even simpler model for
cubic topological insulators.~\cite{Sigrist}, involving 
one Kramers degenerate $d$- and $f$-orbitals. 
These authors have focused on the general topological properties
which are possible in the presence of the cubic symmetry. 
The hopping terms in the tight-binding model ~\cite{Sigrist} involve amplitudes
between nearest-, next-nearest- and next-next-nearest neighbors. As a result, for various choices of the hopping amplitudes various
topologically non-trivial states including the one where $d$- and $f$-bands invert at the X points. Legner et al. have also derived the 
surface state dispersion and found that the Fermi velocity for the electrons is
\beg
v_F=4|V|\sqrt{\frac{|t_ft_d|}{(t_d-t_f)^2}}.
\en
This result implies that the effective mass of the surface electrons $m^*=p_F/v_F$ is quite heavy since the hybridization amplitude is small
compared to other relevant energy scale, while the expression under the square root is of the order $O(1)$.

\section{RECENT EXPERIMENTAL PROGRESS ON \SMB}


Kondo insulators, in particular \SMB{} are
prototypical strongly-correlated mateirals featuring f-electron
physics and some highly unusual electronic properties. As discussed
earlier in this review, the theory of Kondo insulators
has predicted in some KI materials the existence of a topologically
nontrivial surface state with an odd number of Dirac surface bands. In
the past two years tremendous experimental progress has been made on
\SMB{} seeking to identify the existence of the metallic surface
state, to test the topological properties of the surface state and to
search for correlation physics. Below we review some of these
experimental works. We note that despite the rapid progress of the
field, questions remain about the exact nature of the surface state,
the consequence of multiple surface bands, and unusual surface
properties due to interaction within the surface state and between the
surface and bulk Kondo lattice. The surface state of
\SMB{} (and possibly other TKI candidates) a good experimental
platform to explore unusual quantum phases involving strong electron correlations,
heavy fermion physics and topological order.

\subsection{Transport evidence for the metallic surface state}

\begin{figure}
\begin{center}
\includegraphics[width=0.6 \columnwidth] {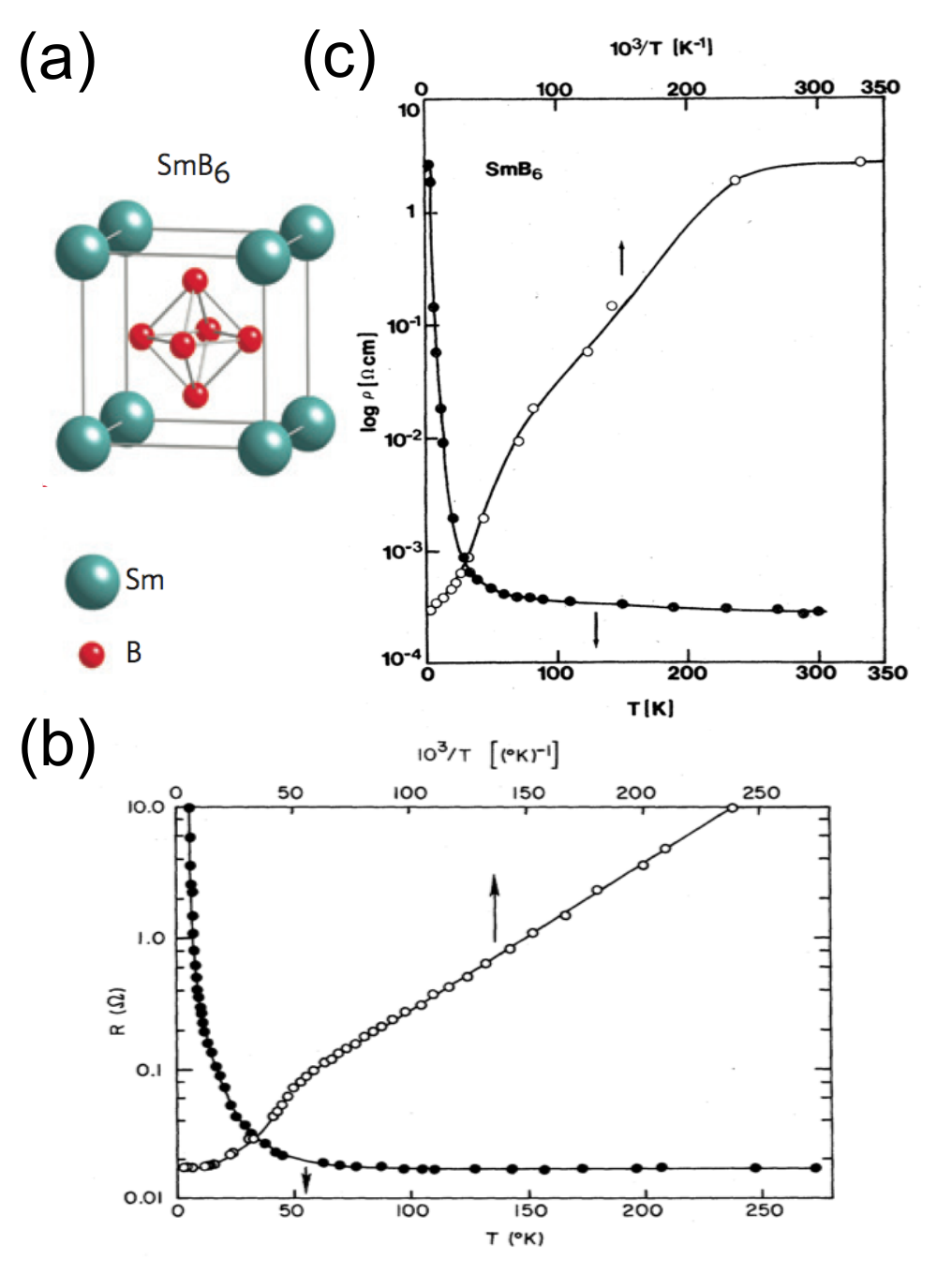}
\end{center}
\caption{\textbf{(a)}, Crystal structure of \SMB{}. \textbf{(b)}, Adapted from ~\cite{smb6}, activated resistance of \SMB6 below 50 K due to the opening of Kondo gap,  \textbf{(c)}, Adapted from ~\cite{allen79}, resistance saturation below 4 K.}
\label{fig:ot}
\end{figure}

\SMB{} (Fig. \ref{fig:ot}a) is a classic Kondo insulator
\cite{Fisk96}. As discussed above, the strong interaction within a
localized periodic dense array of localized f-magnetic-moments leads
to a re-organization of the electronic structure at low temperatures 
\cite{Fisk96} resulting in an energy gap 
driven by hybridization
between conduction electrons and the highly renormalized
f-electrons. When the Fermi level lies within the Kondo energy gap,
an insulator is formed. As a result, \SMB{} behaves as a
correlated metal at high temperatures ~\cite{smb6}, and becomes
insulating below 40 K with the opening of Kondo gap ~\cite{Fisk96},
yielding an activated diverging resistance at low temperatures. This is
illustrated in Fig. \ref{fig:ot}b. However, at even lower temperatures
below 4 K, a peculiar resistance saturation was found in the
original experiments of Menth et al. ~\cite{smb6} and Allen et
al. ~\cite{allen79} (see Fig. \ref{fig:ot}c). This additional
metallic conduction channel was initially suspected to originate from
bulk impurity states ~\cite{smb6}, but the improvement of the sample
quality does not seem to reduce the low temperature saturation
behavior ~\cite{smb6,allen79,Cooley95,Sluchanko00, kimtki}. The theory
of topological Kondo insulator by Dzero et al.~\cite{Dzero2010}
predicted the existence of a topologically protected metallic surface
state (TSS) within this Kondo gap, naturally accounting for the 
resistance saturation. Experiments soon followed to verify
whether the low temperature conduction does indeed occur on the surface.

\begin{figure}
\begin{center}
\includegraphics[width=0.6 \columnwidth] {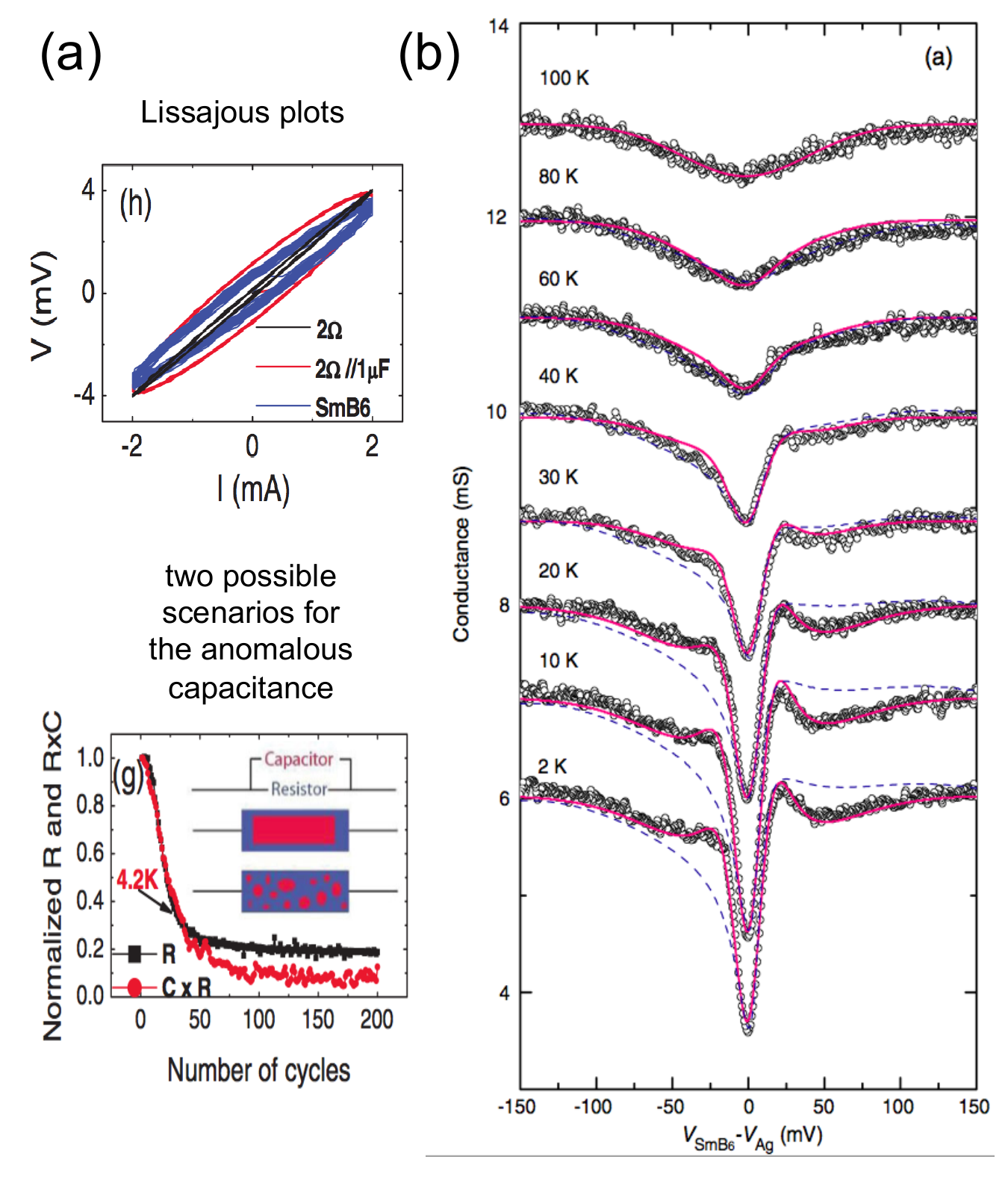}
\end{center}
\caption{\textbf{(a)}, Adapted from Ref.~\cite{kimtki}. From the
Lissajous voltage-current plots \SMB{} is found to behave as a
2$\Omega$ resistor connected in parallel with a 2$\mu$F capacitor. The
anomalous capacitance onsets at 4 K.  \textbf{(b)}, Adapted from
\cite{Zhang13}, temperature dependence of Point-Contact Spectra (PCS)
shows the appearance of a zero bias feature below 100 K, indicating an
opening of the Kondo gap in the bulk. However, the PCS remains the
same as the temperature passes  4K, 
showing that the bulk remains insulating while
additional metallic conductance starts to dominate transport.}
\label{fig:CPCS}
\end{figure}

In 2012, signs of a low-temperature surface state were obtained while
investigating the capacitive self-oscillation effect in \SMB{} single
crystals. Recording the Lissajous plots of current-induced voltage
oscillations, Kim et al., ~\cite{Kim12a} detected an anomalous
capacitance component ($\mu$F) in mm-sized \SMB{} crystals below
4K. This was almost exactly the onset temperature of the saturation
resistance, hinting at a common origin. The response of the \SMB{}
crystal closely resembles that of a 2 $\Omega$ resistor in parallel
with a 2$\mu$F capacitor, as shown in Fig. \ref{fig:CPCS}a. Since a RC
circuit is equivalent electronically to an insulating material
encapsulated with a conducting surface in a bulk material, the surface
conduction picture was a natural explanation for the anomalous
capacitance. However, despite this strong evidence for surface
conduction, alternative scenarios could not be ruled out 
(e.g. having many insulating
grains embedded in a conducting bulk could also lead to an equivalent
RC circuit, as depicted in the inset in Fig. \ref{fig:CPCS}a). This
motivated further transport studies. 

One way to prove that \SMB{} is truly insulating in its bulk 
is to
show that the bulk density of states (DOS) remains unchanged 
below the resistance saturation temperature. 
This
can be done by measuring the resistance of a very fine contact to the
sample: a technique called {\sl Point-Contact Spectroslcopy} (PCS). If the
low temperature residual conduction arises from bulk impurity state or
the above-hypothesized metal-insulator mixture instead of a metallic
surface, the bulk electronic state would necessarily change when the
coherent conduction channel develops at low temperatures. Zhang et
al. ~\cite{Zhang13} performed PCS on \SMB{} using Ag particle point
contacts. They recorded the point-contact conductance as functions of
both Ag-\SMB{} voltage difference and temperature
(Fig. \ref{fig:CPCS}b). The former conductance-voltage relation, or
conductance spectra, give valuable information regarding the bulk
electronic structure at a given temperature. Zhang et
al. ~\cite{Zhang13} observed the onset of a zero-bias conductance dip
below 100K, signaling the opening of the Kondo gap due to electron
correlations. Perhaps the most striking finding of the experiment is
that the conductance spectra remained unchanged while the temperature
was lowered from 10 K to 2K, where the sample resistance saturation
occurred. This proves directly that during the resistance saturation,
the bulk of \SMB{} remains insulating with an almost constant Kondo
gap. The most natural explanation of this phenomenon is to
assign the low-temperature metallic conduction to the surface.

\begin{figure}
\begin{center}
\includegraphics[width=0.6 \columnwidth] {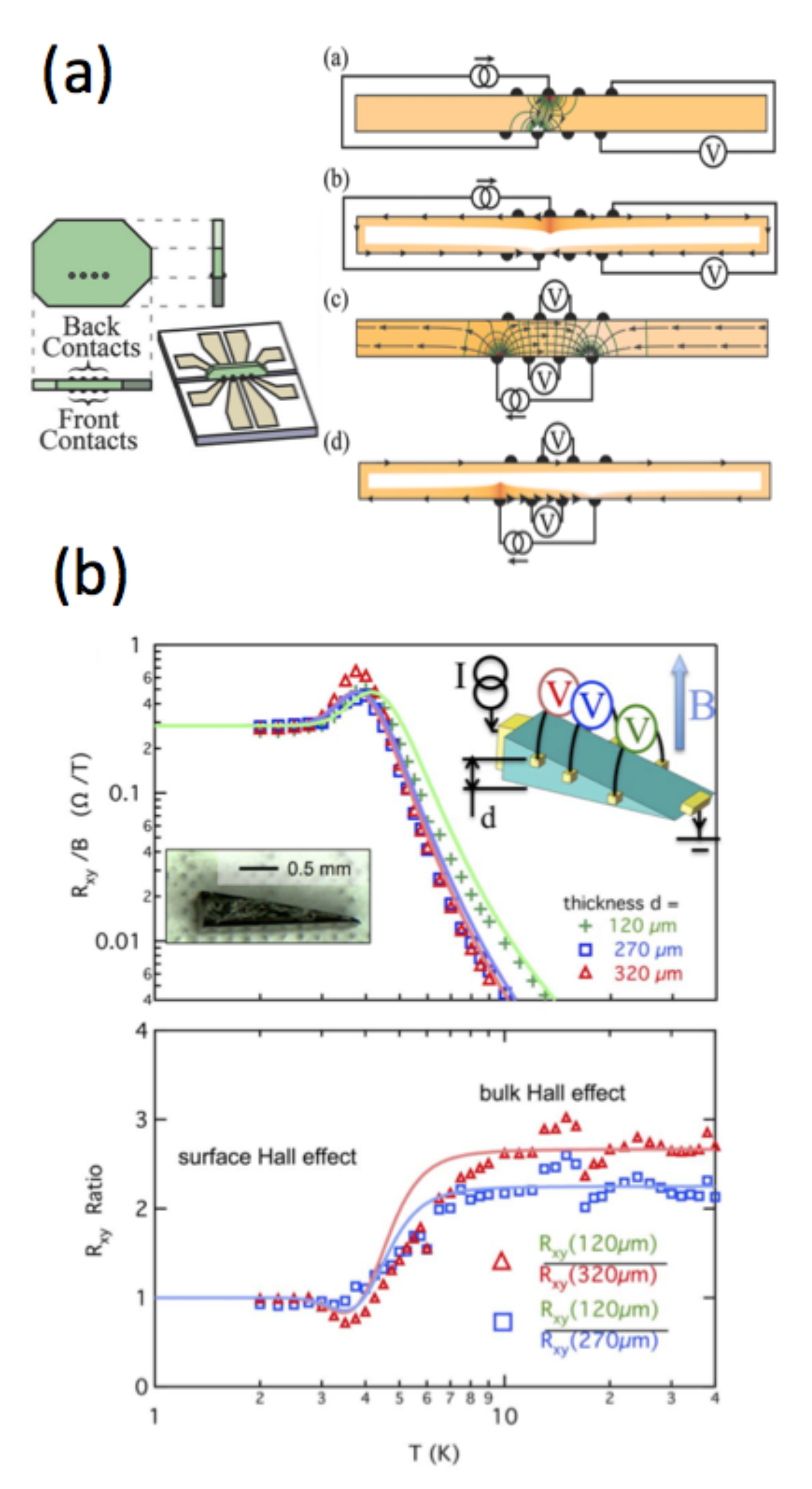}
\end{center}
\caption{\textbf{(a)}, Adapted from Ref.~\cite{wolgasttki}, transport measurement with various current-voltage configurations on a thin plate \SMB{} crystal. Depending on bulk or surface conduction, the relative size of the voltages would be significantly different. \textbf{(b)}, Adapted from ~\cite{kimtki}, Hall effect measurements at different thicknesses in a single wedge-shaped \SMB{} Hall bar. The Hall voltages were found to be inversely proportional to thickness at high temperatures, indicating bulk conduction. Below 4K, the Hall voltage becomes thickness independent, signaling 2-dimensional surface conduction at low temperatures.}
\label{fig:nlHall}
\end{figure}

The low-temperature surface conduction could also be demonstrated somewhat more directly by noticing that the bulk and surface contributions to conduction vary differently as one changes the size and shape of the sample. Wolgast et al. ~\cite{wolgasttki} designed and fabricated a thin-plate-shaped \SMB{} crystal (Fig.~\ref{fig:nlHall}a) in order to enhance the difference between bulk and surface contributions. The electric current leads were placed on the top and bottom sides of the sample, which was sandwiched between two insulating silicon pieces. For the bulk conduction in this \SMB{} plate, the current can zip through the middle thin part of the crystal; while the surface current has to travel along the surface around the corners of the sample. Therefore, voltage leads, placed at various locations on the sample, could register these two vastly different current paths. In their experiments ~\cite{wolgasttki} three voltages (Fig.~\ref{fig:nlHall}a) are measured as a function of temperature. After comparison between theoretical simulations and experimental results, it was found that surface conduction indeed dominated at low temperatures, while bulk contribution dominated at high temperatures ~\cite{wolgasttki}.

Another transport quantity that would differ sharply between surface
and bulk conduction mechanisms is the Hall effect. In a
perpendicular magnetic field, the transverse or Hall voltage is
inversely proportional to the thickness of the Hall bar in the case of
the usual bulk conduction. However, if surface conduction dominates,
the thickness of the Hall bar is irrelevant and the Hall
voltage would be thickness-independent. Therefore the
thickness-dependence of the Hall effect should directly indicate whether
bulk or surface conduction dominates. A technical difficulty
associated with such an experiment is the innevitable variability in crystal
thicknesses.

This problem
was solved in an experiment by Kim and Thomas {\em et al.}~\cite{kimtki}
who measured Hall voltages at different thicknesses, within the same
crystal using an \SMB{} Hall bar shaped into
a thin and long wedge (Fig.~\ref{fig:nlHall}b).
In smooth, well-polished \SMB{} wedges, they found
that the
high temperature Hall resistance \Rxy{} = \Vxy{}/I was inversely
proportional to the thickness, consistent with bulk transport ~\cite{kimtki} . 
The temperature
dependence (Fig. ~\ref{fig:nlHall}b) of \Rxy{} follows an activated behavior
with a transport activation gap of 38K, confirming the insulating
nature of the bulk. However, at temperatures below 4 K, the Hall
resistance \Rxy{} becomes  thickness-independent, proving
the domination of surface conduction. At these temperatures, the
\Rxy{} is also temperature independent, consistent with the metallic
nature of the surface conduction. Without a magnetic field, the
qualitative evidence for low temperature surface conduction was also
obtained in Ref.~\cite{kimtki} by comparing the local and non-local
voltages on thin-plate samples, as was used in non-local voltage
measurements ~\cite{McEuen90} in quantum Hall effect systems to
identify the then-debated edge state conduction. This measurement is
conceptually equivalent to the thin-plate measurement described above
\cite{wolgasttki}, both taking advantage of the vastly different paths
of surface and bulk conductions. Several samples were used for both
Hall and non-local measurements, including purposely scrapped
samples. And the surface dominated conduction was found to hold
independent of the crystal surface direction or surface quality, which
is consistent with a topologically protected surface state that can
only be destroyed by broken-time-reversal, e.g. magnetic,
perturbations.
 
Perhaps an even more vivid demonstration of the low temperature
surface conduction is to show that the longitudinal resistance doesn't
depend on the thickness of thin samples. In such an experiment, 
special care needs to be
taken 
to preserve the surface
quality before and after the thinning process. Nevertheless positive
results have been demonstrated by several groups now
\cite{Kim12a,wolgasttki,Zhang13,Phelan14}.

\subsection{ARPES studies of the surface state}\label{arpes_studies}

\begin{figure}
\begin{center}
\includegraphics[width=0.6 \columnwidth] {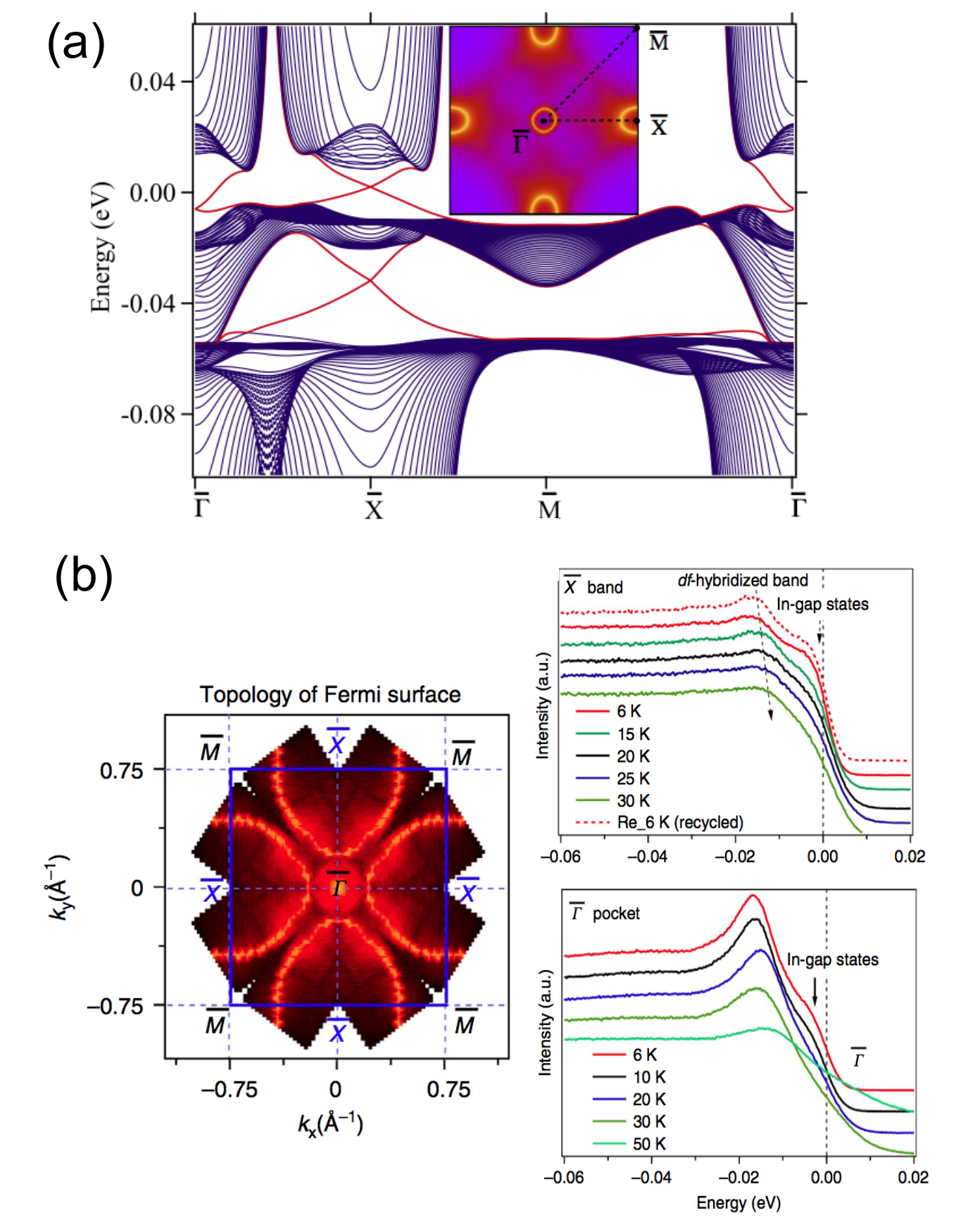}
\end{center}
\caption{\textbf{(a)}, Adapted from Ref.~\cite{Lu13}, calculated electronic structure of \SMB{} for the first Brillouin zone of the (001) surface (inset). The blue lines are bulk energy levels and the red lines are surface states, showing three Direct cones.  \textbf{(b)}, Adapted from ~\cite{Neupane13}, topology of the Fermi surface measured by ARPES, consistent qualitatively with the theoretical calculation ~\cite{Lu13}. The left figure is a temperature evolution of the ARPES spectral intensity, showing the onset of the in-gap (surface) state at low temperatures.}
\label{fig:arpes1}
\end{figure}

The electronic structure of the surface state can be probed directly
by angle-resolved photoemission spectroscopy (ARPES). An observation
of odd number of Dirac surface bands would be a direct confirmation of
the topological nature of the surface state. In addition,
spin-resolved ARPES can in principle investigate the spin structure of
the surface state. This has led to a flurry of recent ARPES studies of
\SMB{}.  Because of the limited energy resolution of ARPES relative to
the small Kondo gap of \SMB{}, the detailed spin structure of the
surface states is currently still
contraversial\cite{Xu:1jv,Hlawenka15}.  Important insights regarding
the energetics of the surface state have been nevertheless been
obtained by several recent ARPES studies. Before we review these
experimental results, it would be helpful to first take a look at the
predicted \SMB{} energy spectrum for a topological Kondo insulator 
obtained by Lu et al. ~\cite{Lu13}. Another calculation has been
carried out by Alexandrov et al. ~\cite{Alexandrov13}. One calculated
energy spectrum \cite{Lu13} is shown in Fig. \ref{fig:arpes1}a for the
first Brillouin zone (BZ) in (001) surface, where high-symmetry points
are labeled as $\Gamma$, X, and M. The blue lines are bulk energy
levels with a Kondo gap at the Fermi level. And the red lines are
surface states, showing three Direct cones at $\Gamma$, and the two X
points.

\begin{figure}
\begin{center}
\includegraphics[width=0.6 \columnwidth] {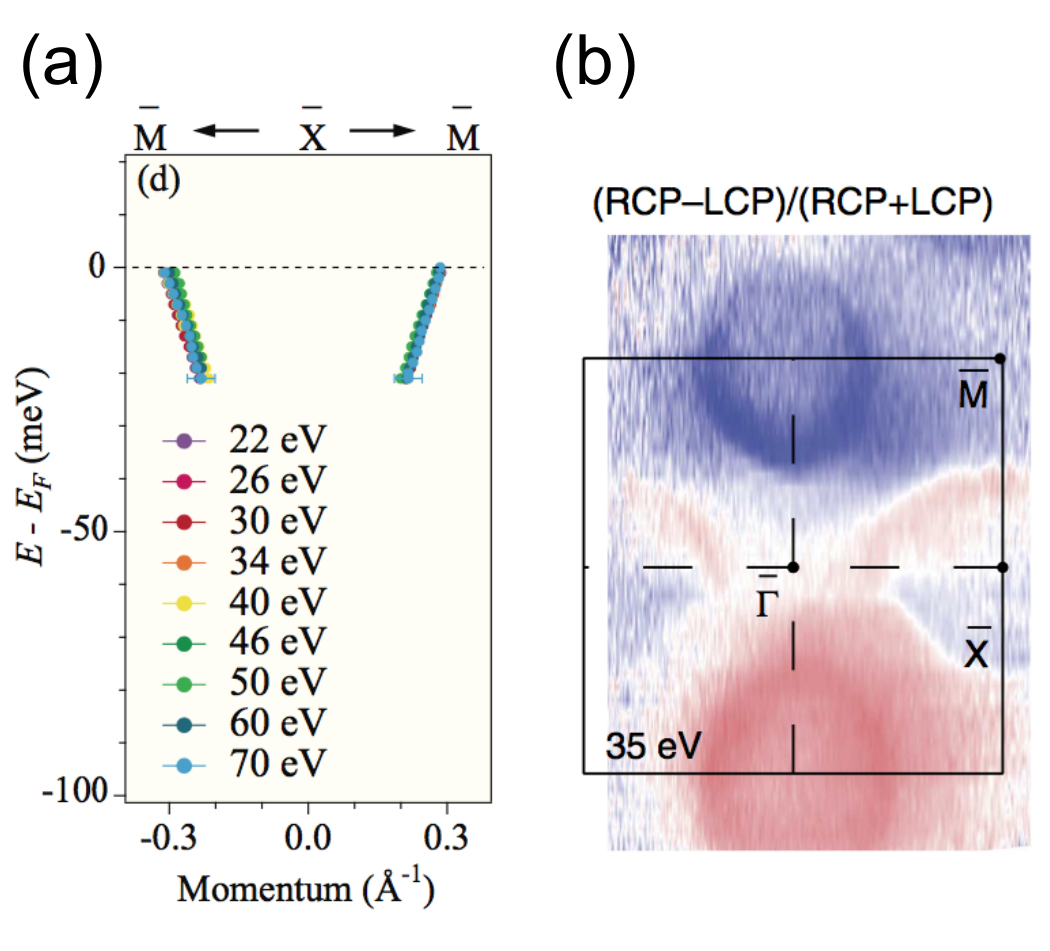}
\end{center}
\caption{\textbf{(a)}, Adapted from Ref.~\cite{Xu13}, extracted dispersions of the $\beta$ band (X pocket) for different photon energies. The linear dispersions demonstrate the two-dimensional nature of the pocket.  \textbf{(b)}, Adapted from ~\cite{Jiang13}, the differential map of the photoemission intensities with left-circularly-polarized and right-circularly-polarized  light. The anti-symmetric pattern resembles that of a topological surface state.}
\label{fig:arpes2}
\end{figure}

The ARPES results of Neupane et al. ~\cite{Neupane13} are replotted in
Fig. \ref{fig:arpes1}b for direct comparison. The topology of the
measured Fermi surface at 6 K agrees well with the topological Kondo
insulator calculation (Fig. \ref{fig:arpes1}a). As predicted, three
low-lying metallic states near the Fermi energy exist at the $\Gamma$,
and the two X points. The two X pockets are quite large,
suggesting that the Fermi energy is far away from the (possible) Dirac
points at X points. Two other important results are obtained by
Neupane et al. ~\cite{Neupane13} from the temperature evolution of the
ARPES spectral intensity (Fig. \ref{fig:arpes1}b). The first is that
the in-gap metallic state only emerges at temperatures below 15 K for
both $\Gamma$ and X pockets, suggesting the nontrivial origin of the
metallic state. The second finding is that the metallic in-gap state
does not disappear after a 6 to 50 K thermal recycling, which
demonstrates that the in-gap states are robust and protected against
thermal recycling.

Two additional ARPES results by Xu et al. ~\cite{Xu13} and Jiang et
al. ~\cite{Jiang13} were reported at the same time which 
are consistent
with the topological Kondo insulator scenario for SmB$_{6}$.
In particular in
Ref.~\cite{Xu13}, Xu et al. extracted the energy dispersions of the
$\beta$ band (X pocket) (Fig. \ref{fig:arpes2}a). The linear
dispersions demonstrate the two-dimensional nature of the
pocket. Unfortunately the Dirac points could not be observed clearly
in this study. For the $\beta$ band, Xu et al. found that the
intensity diminishes suddenly at 20 meV below Fermi level,
corresponding to the hybridization gap edge between f and d
electrons. They argue that this may prohibit observing the Dirac point
formed by the bands crossing each other. In ~\cite{Jiang13} Jiang et
al. performed circular dichorism (CD) ARPES measurements.  CD ARPES
may shed some light on the spin structure of the metallic state.
Jiang et al. calculated the differential map of the photoemission
intensities with right-circularly-polarized (RCP) and
left-circularly-polarized (LCP) lights as repotted in
Fig.~\ref{fig:arpes2}b. The anti-symmetric pattern with respect to the
$\Gamma$-X line resembles that of a topological surface
state. However, it is not yet sufficient to make conclusive statements
about the surface spin texture or Berry's phase ~\cite{Neupane13} due
to several complexities related to CD ARPES.

Parallel ARPES studies on \SMB{} were carried out by Zhu et
al. ~\cite{Zhu13} and Denlinger et al. ~\cite{Denlinger13}
respectively. In Ref.~\cite{Zhu13}, Zhu et al. propose an alternative
explanation for the conducting surface state, suggesting that it 
originates from boron dangling bonds on the (001)
crystal surface, i.e. that it is a non-topological polar surface. 
In \cite{Denlinger13} Denlinger et al. found high ARPES
intensity at the H points, where  the momentum-location of the H-point along
the $\Gamma$-M direction coincided with the polar metallic surface
state claimed by Zhu et al. ~\cite{Zhu13}.  However, Denlinger et
al. ~\cite{Denlinger13}, found that the H-point is  gapped at low T
temperature, suggesting that the polar surface state found by Zhu et
al. ~\cite{Zhu13} is in fact insulating and hence unrelated to the
metallic surface state. Due to the limited resolution, it is somewhat
hard to judge which scenario is correct at the moment. However,
the (001) polar metallic surface picture by Zhu et al. ~\cite{Zhu13}
seems to contradict available transport experiments that showed the
existence of surface metallic conduction on both (001) and (011)
surfaces. The latter surface is not polar and therefore can not host a
polar metallic surface state.

\begin{figure}
\begin{center}
\includegraphics[width=0.6 \columnwidth] {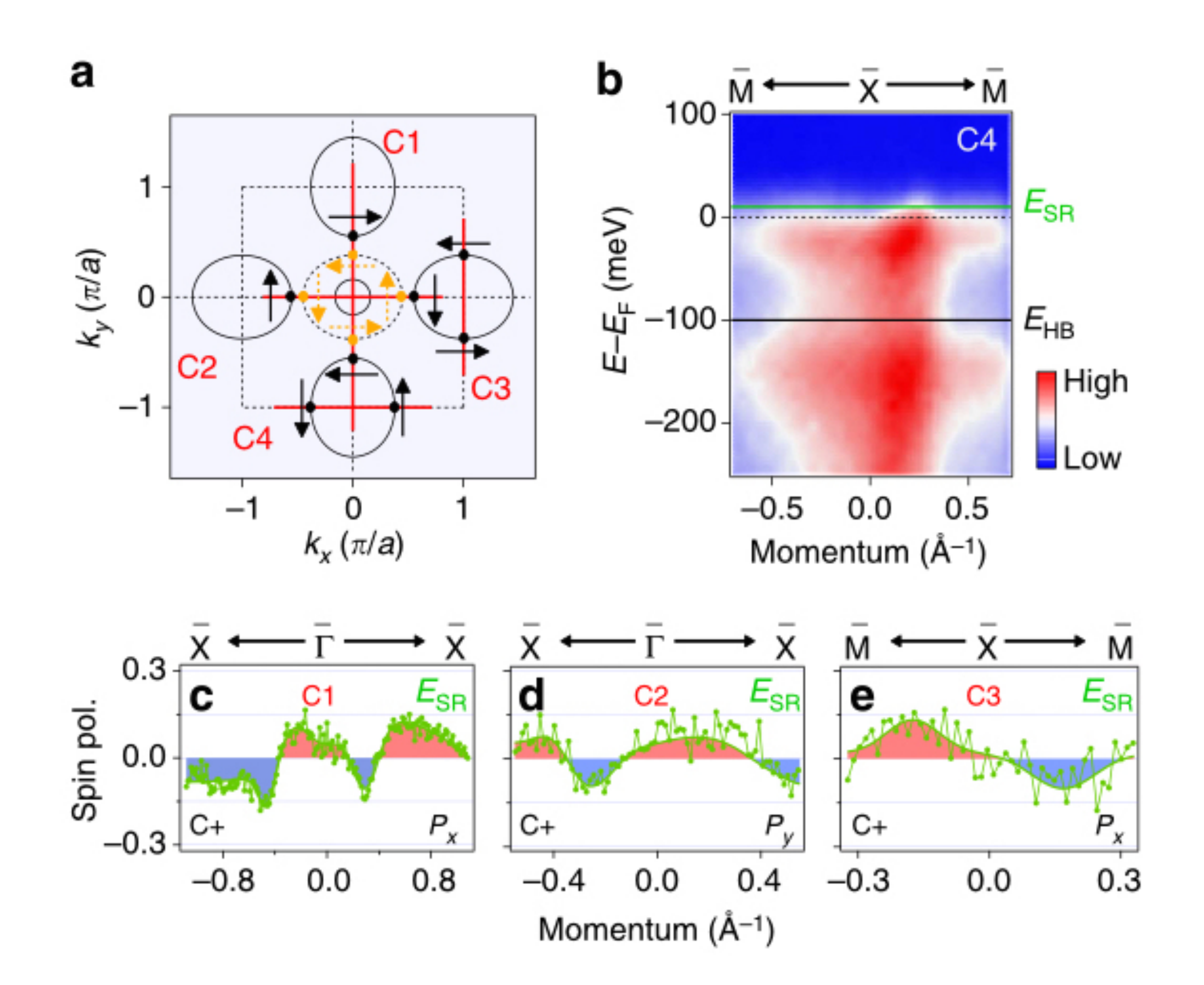}
\end{center}
\caption{\textbf{(a)} Adapted from Ref.~\cite{Xu:1jv}, 
showing the spin polarization of the surface states observed along the
high-symmetry lines. The red lines indicate the locations of spin
measurements in the $k_{x}$, $k_{y}$ plane, labelled as C1, C2, C3 and
C4. The arrows indicate the measured spin polarizations at the
positions of the black dots. In  \textbf{(b)}. The low-energy excitations
along the high-symmetry line $\bar M-\bar  X-\bar M$. \textbf{(c)- (e)}
show the spin polarization measured at 30, 30 and 26 $eV$ respectively
at points C1, C2
and C3, with the spin polarization along the x, y and x directions
respectively. 
}
\label{SARPES}
\end{figure} 

Finally we mention the current status of spin-polarized ARPES
measurements.  In weakly interacting topological insulators, the
classic test for topological Dirac states, is to directly observe
their spin texture using spin-resolved ARPES
measurements\cite{hasanclassic}.  However, in SmB$_{6}$, the
narrowness of the bulk insulating gap makes this a highly challenging
measurement, the results of which are still contraversial.  The first
spin-polarized ARPES measurements have been carried out on SmB$_{6}$
were carried out at the Swiss Light Source\cite{Xu:1jv}. These
measurements show of spin-momentum locked states on the 100 surfaces
in the vicinity of the $\bar X$ points in the surface Brillouin zone
(see Fig. \ref{SARPES}), supporting the presence of topological Dirac
states.  However, spin-polarized ARPES measurements carried out at the
Helmholtz-Zentrum, Berlin\cite{Hlawenka15} have led this group to
question the topological interpretation. In particular, their
measurements around the $\bar \Gamma$ points appears to support the
presence of Rashba-split, conventional surface states of a trivial
insulator.  The resolution of this experimental controversy awaits
improved resolution and further measurements.

\subsection{STM studies of the surface state}

\begin{figure}
\begin{center}
\includegraphics[width=0.6 \columnwidth] {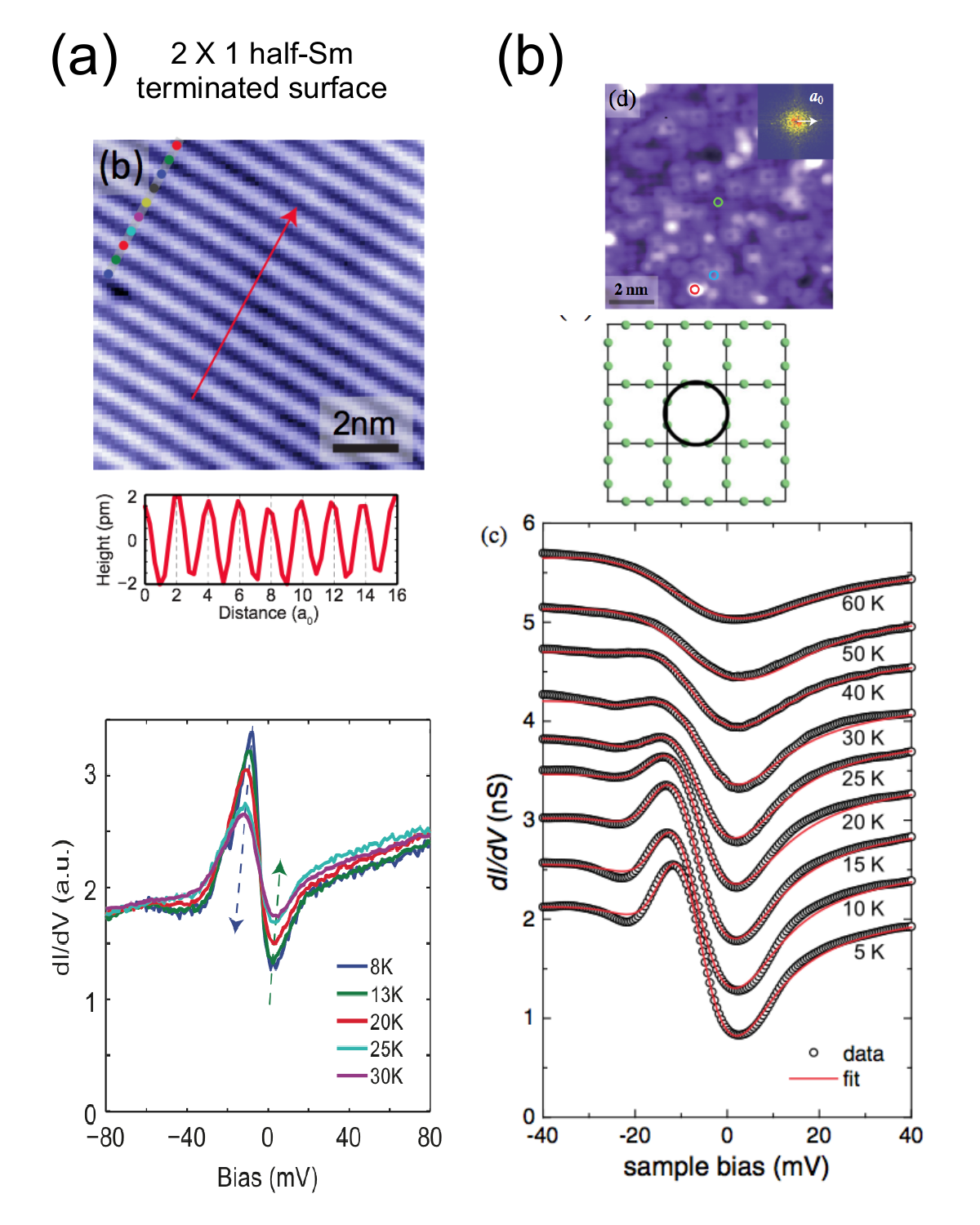}
\end{center}
\caption{\textbf{(a)}, Adapted from Ref.~\cite{Yee13}, STM morphology image of a $2 \times 1$ half-Sm terminated surface and its STS spectra for various temperatures.  \textbf{(b)}, Adapted from Ref.~\cite{Ruan14}, STM morphology image of one of the four phases found by the STM study of Duan et al. The temperature dependence of STS spectra for the ``donut'' D2 phase.}
\label{fig:stm}
\end{figure} 

Scanning Tunneling Microscopy (STM), is another 
useful surface probe with the potential to reveal the properties of the
surface states with atomic spatial resolution. Spectroscopic
measurement of the tunneling conductance as a function of bias
voltage, or Scanning Tunneling Spectroscopy (STM) further reveals the
local density of states near the Fermi energy. Therefore a
temperature-dependent STS study would tell quite a lot about the
energetics of bulk and surface state, complementary to ARPES
measurements.

STM and STS study on the (001) surface \SMB{} by Yee et
al. ~\cite{Yee13} revealed that the surface of \SMB{} is quite
complex and far from an ideal. 
Their STS morphology image of the normal (001) surfaces
shows that much of the surface 
reconstructs to a $1 \times 2$ structure in which half the surface is
samarium-terminated (See: Fig. \ref{fig:stm}a). Other regions of the surface appear
amorphous and only small regions involve a pristine termination. 
It is unclear at this stage if the
reconstruction is due to the particular cleavage method or whether it is
intrinsic to \SMB{}. Nevertheless, this STM study suggests that the
ARPES studies, which have optical spot sizes of hundreds of microns
could have picked up a mixture of spectra from a variety of 
different surface morphologies.  This may explain in part the
limited resolution of \SMB{} ARPES measurements.

In the STM study by Yee et al. ~\cite{Yee13} a common STS spectra
feature was observed for both $1\times 1$ and $1 \times 2$
regions: a minimum
of dI/dV at the Fermi level and a gap (dI/dV peak) slightly below this
minimum in energy. Fig. \ref{fig:stm}a shows the STS spectra for the
case of $1\times 2$ region, where the peak is centered at - 8 meV and
becomes smaller at higher temperatures. Extrapolating to high
temperature, Yee et al. argued that the peak (gap) would vanish at
40K, suggesting its hybridization nature. They also observed residual
spectral weight spanning the hybridization gap down to the lowest
temperatures, which is consistent with a topological surface
state. Qualitatively similar findings were observed for the $1\times 1$
surface region, although the energy is shifted by tens of eV. This
testifies the robustness of the hybridization gap. In the other STM
work ~\cite{Ruan14} by Ruan et al., four types of surface morphology
were found, highlighting again the complexity of the \SMB{}
surface. The STM image for the ``Donut'' D2 phase is shown in
Fig. \ref{fig:stm}b with a cartoon illustrating the ``Donut''
structure. The temperature dependence of the STS spectra
(Fig. \ref{fig:stm}b) is qualitatively similar to those found by Yee et
al. 
The data revealed the emergence of
a resonance peak below 40K, signaling the opening of the Kondo
gap. Summarizing the body of all STM works, the large zero bias conductance
that persists to the lowest temperature may come from the topological
surface state. However a smoking gun STM evidence for TKI is still
lacking: the quasiparticle interference patterns showing surface state
with helical spin texture.

\subsection{Is the surface state topological?}

\begin{figure}
\begin{center}
\includegraphics[width=0.6 \columnwidth] {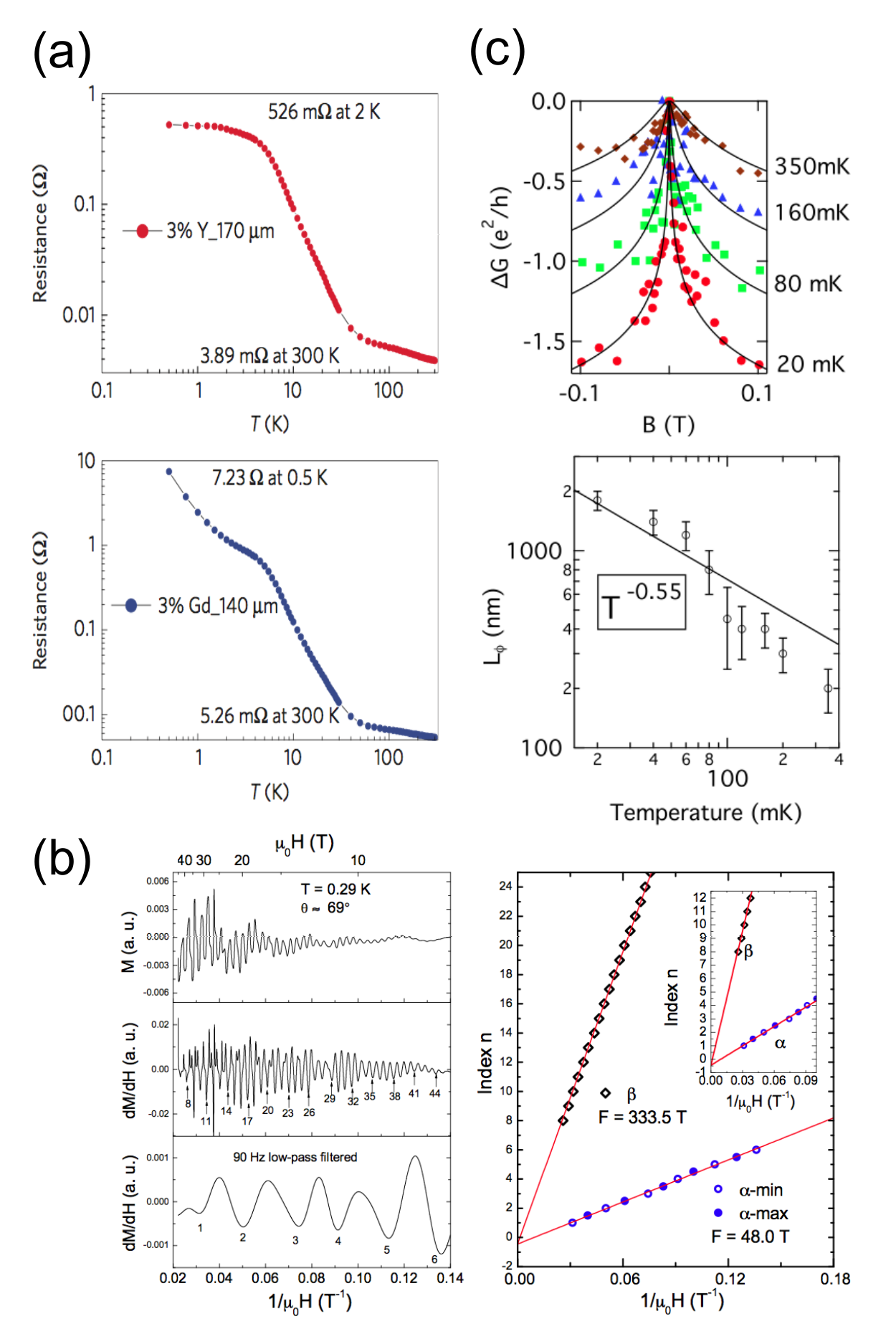}
\end{center}
\caption{\textbf{(a)}, Adapted from Ref.~\cite{Kim13}, contrasting behavior between non-magnetic Y-doped and magnetic Gd-doped \SMB{} samples. Magnetic doping destroys the metallic surface state.  \textbf{(b)}, Adapted from Ref~\cite{Li13}, de Haas van Alphen quantum oscillation from a \SMB{} sample. Extrapolation of the Landau level index may indicate a Berry phase of $\pi$. \textbf{(c)}, Adapted from Ref.~\cite{Thomas13}, weak anti-localization effect in \SMB{} as expected from a topological surface state. }
\label{fig:topo}
\end{figure} 

The experiments reviewed up to this point, provide strong evidence for the
existence of the metallic surface state that emerges from an
insulating bulk state at low temperatures. The metallic surface state
could be conveniently probed by transport methods due to the
insulating nature of the bulk. Although the low temperature emergence
of the surface state ~\cite{Neupane13} and the antisymmetric ARPES
pattern with opposite chirality of light ~\cite{Jiang13} seem to be
consistent with a topological surface state, more direct evidence is
still needed to tell whether the surface state is really
topological. In an ideal world, energy- and spin-resolved ARPES conducted at low
temperatures below the resistance saturation would confirm whether 
the spin texture of the surface state has a Berry phase. 
However, the required high energy resolution (few meV) is
beyond the current capability of both synchrotron ARPES (40 meV) and
laser ARPES (10 meV). Other difficulties derive from the limited low
temperature range and spin resolution with spin-resolved ARPES
techniques.  It is noted that other TKI candidates may possess larger
Kondo gaps, making them to be within the resolution of current
spin-resolved ARPES capabilities. As mentioned above a smoking gun STM
evidence for the topologic surface state would be the quasiparticle
interference patterns showing surface state with helical spin
texture. This might be within the capability of current STM
techniques, if a better surface without reconstruction could be
prepared for STM studies.

Since the bulk of \SMB{} is insulating at low temperatures, it is
possible to test the topological properties of the surface state,
giving partial evidence for the topological surface state. A
topological surface state has three aspects of topological
protection. First, their fundamental Z2 topology preserves a gapless
surface state unless time reversal symmetry (TRS) is broken. Second,
helical spin polarization prevents momentum backscattering from $-k$
to k by non-magnetic impurities. Finally, the Berry phase protects the
surface state from weak localization through time reversed
paths. These collectively provide a robust surface state with TRS
conservation. Therefore for a topological surface state, it could only
be made insulating by broken TRS such as those created by 
magnetic perturbations. In
addition, such a surface state would become less conductive when a
small magnetic field is applied, destroying the negative interference
between time-reversed carrier paths. This is called the
weak-antilocalization (WAL) effect ~\cite{Altshuler85}. These two
experimental tests have been carried out by Kim et al. ~\cite{Kim13}
and Thomas et al. ~\cite{Thomas13} and both have yielded positive results in
agreement with a topological surface state.

Kim et al. ~\cite{Kim13} made a comparison study between \SMB{}
crystals with magnetic (Gd) impurity doping and non-magnetic (Y or Yb)
impurity dopings. In those thin-plate shaped samples, surface vs bulk
conduction could be distinguished by performing thickness dependence
measurements: when surface conduction dominates, the resistance ratio
should be irrelevant to thickness. Kim et al. found that Yb and Y
doped \SMB{} exhibit bulk conduction at high temperature and change to
surface conduction at low temperatures, similar to pure \SMB. In
contrast, the resistance of Gd doped \SMB{} remains inversely
proportional to sample thickness for all temperatures, indicating
bulk-dominated conduction even at low temperature. A perhaps more
direct illustration of the contrasting impact from magnetic and
non-magnetic dopings is plotted in Fig. \ref{fig:stm}a. Below 4 K, the
resistance of the Y doped sample saturates just like in pure \SMB{},
while the Gd doped sample shows insulating behavior. Only the broken
TRS magnetic doping destroys the metallic surface state. It is noted
that in this study, the amount (3 percent) of Gd doping is
sufficiently small to not introduce order magnetic states, as
evidenced from the magnetic measurements ~\cite{Kim13}.

Weak antilocalization (WAL) ~\cite{Altshuler85} is expected in a
topological surface state due to an unusual Berry phase of $\pi$
\cite{Qi11,Hasan10}, which causes destructive interference between
time-reversed electron paths and lowers the sample resistance. This
effect is destroyed by a time-reversal-symmetry-breaking magnetic
field, giving rise to magneto-resistance dip around zero
field. Unfortunately, 
WAL alone can not provide conclusive evidence of a
topological surface state because strong spin-orbit coupling can also 
produce a WAL effect. Indeed Thomas et al. observed clear WAL
effect in \SMB{} samples at low temperatures ~\cite{Thomas13}. As
shown in Fig. \ref{fig:stm}c, the WAL manifests itself as a
conductance peak at zero field, with the height of the peak decreases
at higher temperatures. The shape of the WAL feature fits well to the
Hikami-Larkin-Nagaoka (HLN) equation ~\cite{Hikami80}. The temperature
dependence of the coherence length could be extracted using the
fittings, suggesting that below 80 mK, electron-electron interaction
dominates the scattering of transport current. The WAL effect has also
been detected by Nakajima et al. ~\cite{Nakajima13}.

The topological aspect of the surface state could also be probed by
quantum oscillations: the periodic variation of physical properties of
the sample as the Landau levels pass through the Fermi level in a
changing magnetic field. For a topological surface state with odd
number of Dirac points, the Berry's phase would be $\pi$ instead of
the usual value of 0. As a result, the Landau level index would be
half-integer at the large field limit. Using a sensitive torque
magnetometry setup, Li et al. ~\cite{Li13} have recently reported 
the detection of 
de Haas van Alphen oscillations (dHvA) 
in the magnetization of \SMB crystals. 
Fig. \ref{fig:stm}b shows two types of dHvA
oscillations in that sample, identified as the 
(001) and (011) surfaces respectively. The Landau level index of both
oscillations extrapolate to -1/2 at the high field limit, suggesting
a non-trial Berry 
phase of $\pi$ in both energy bands. 
Their experiment finds that the frequencies of the observed
oscillations only  depend on the perpendicular component of the
field, as expected for a surface state.  One of the intriguing aspects
of the measurements, is that the quasiparticle effective mass
obtained the temperature dependence of the signal is about 0.1m$_{e}$.
This small mass is unexpected on the basis of the ``Heavy
Fermion'' nature of \SMB. More experimental and theoretical
investigations are needed to understand this discrepancy.

\begin{figure}
\begin{center}
\includegraphics[width=0.6 \columnwidth] {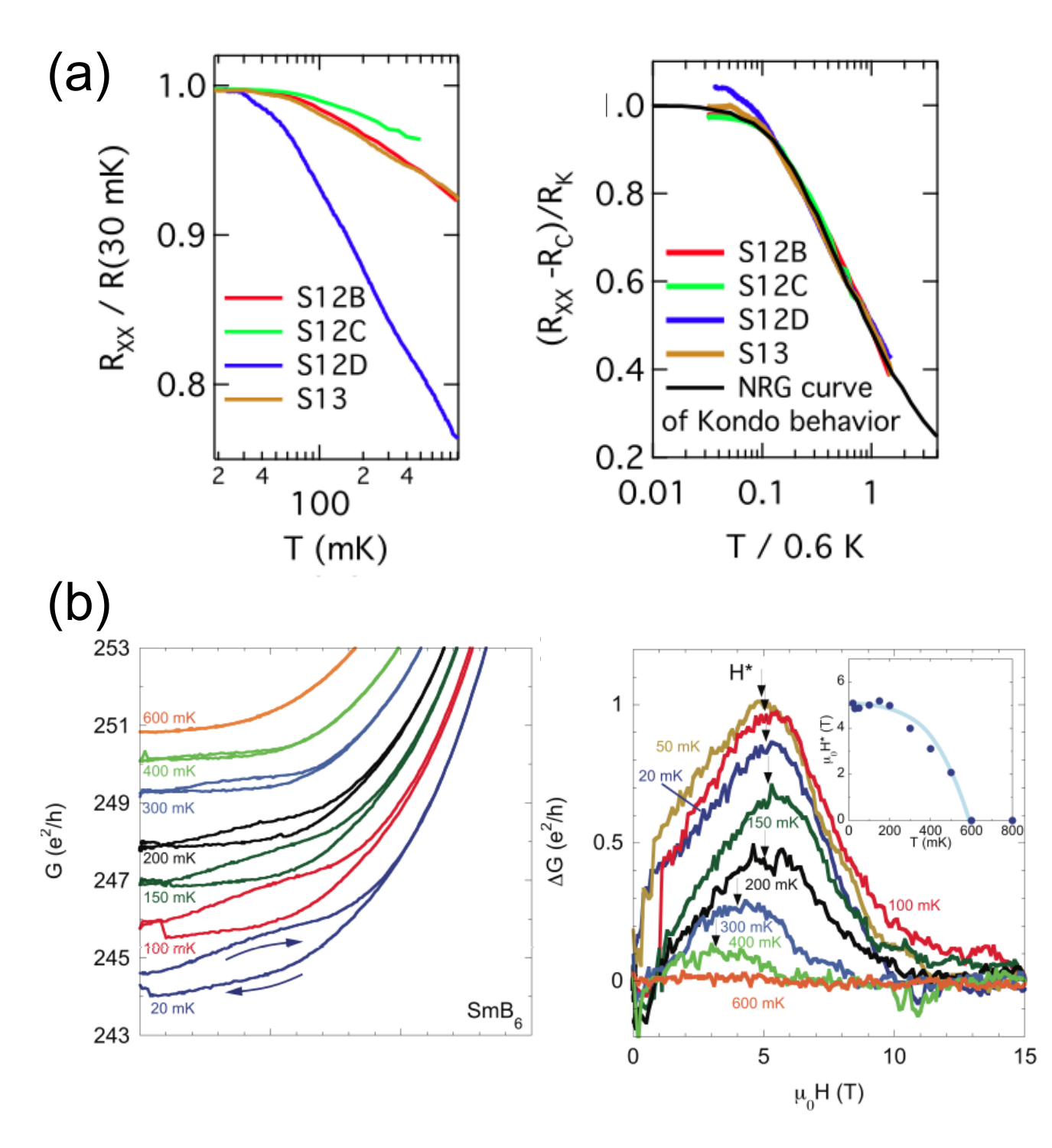}
\end{center}
\caption{\textbf{(a)}, Adapted from Ref.~\cite{Thomas13}, very low temperature resistance saturation of several \SMB{} samples, showing $\log(T)$ divergence before saturation. \textbf{(b)}, Adapted from Ref.~\cite{Nakajima13}, hysteretic magnetoresistance. The difference of conductance between hysteresis lie within the range of quantum conductance. This could be explained by one-dimensional edge states one the surface between spin polarized domains. }
\label{fig:kondo}
\end{figure} 
\section{OPEN QUESTIONS}

Topological Kondo insulators represent a collision of three
areas of research: topological matter, strongly correlated systems and
surface science, and experts from each of these fields find themselves
outside their comfort zone. It is against this backdrop that 
\SMB{} as a candidate TKI poses a 
immense new theoretical, experimental, and materials science challenges. 

The physics of Kondo insulator \SMB{} originates from a periodic dense
lattice (Kondo lattice) ~\cite{Fisk96} of localized magnetic moments,
and is quite different from the Kondo effect ~\cite{{Kondo64}} of 
isolated magnetic moments. Moreover, while it may be adiabatically connected to
a topological band insulator, the system also lies at the brink of
magnetism. Indeed, 
high pressure experiments show that at pressures of about 9GPa, the
hybridization gap of SmB$_6$ closes, leading to the development of 
magnetic order\cite{hpmag1,hpmag2}.
This raises many interesting questions and possibilities, for instance:
\begin{itemize}
\item
What is the effect of the surface on the Kondo effect?
\item Could the bulk and its excitations
be different in any way from a conventional insulator?
\item Will magnetism emerge on the surface, and if so, 
what kinds of highly correlated and competing order 
are realized near the point of instability?  
\item Finally, what is the effect of disorder, such as vacancies or ``Kondo
holes'' at the Samarium sites? 
\end{itemize}
In a study by Thomas et al. ~\cite{Thomas13} who 
followed the saturated resistivity 
of \SMB{} to mK temperatures, the resistivity 
was found to exhibit a $log(T)$ temperature dependence 
(Fig. \ref{fig:kondo}a) before the final saturation below 40 mK. The
size of the $log(T)$ behavior is both sample-dependent and
inconsistent with the WAL effect alone even with the correction of
electron correlation. However, the temperature dependence can be
perfectly explained by assuming the Kondo effect in the surface state
driven by by Kondo holes, with a universal Kondo temperature of 0.6 K. In fact,
the data from several samples could be scaled into a single numerical
renormalization group calculation of Kondo effect behavior assuming a
single Kondo temperature (Fig. \ref{fig:kondo}a).

Nakajima et al. reported hysteresis in magneto-resistance
\cite{Nakajima13} in a \SMB{} sample, as shown in
Fig. \ref{fig:kondo}b. The hysteresis may originate from ferromagnetic
state on the surface, which they hypothesize to arise due to Kondo
holes. The edge state between magnetic domains of opposite chirality
could in principle yield quantized (Hall) conductance. Nakajima et
al. observed that the hysteretic part of the longitudinal conductance
lies within the range of the quantum conductance
(Fig. \ref{fig:kondo}b), hinting the aforementioned mechanism.

It is clear from these examples, that 
the effects of interactions on
topological surface states pose a major challenge.
There have been a number of recent
theoretical publications that have started to explore these issues
\cite{altshuleraleiner,BitanRoy,Erten15,galitski2}. 
Bitan Roy and collaborators\cite{BitanRoy} have emphasized an
itinerant description of the interacting surface Dirac cones and
in\cite{galitski2} 
predict a proclivity towards excitonic and nematic instabilities.
The papers \cite{altshuleraleiner,Erten15} discuss the physics in
terms of a surface Kondo lattice involving
local moments interacting with chiral surface states.
Alshuler and Aleiner have considered a simplified one-dimensional surface
state, and show that interactions with the local moments will 
localize surface states. Alexandrov, Erten and Coleman\cite{Erten15}
have proposed a similar two dimensional model and propose that the
surfaces of topological Kondo insulators may develop quantum critical
ground-states. 

One of the outstanding puzzles 
concerns the very high group velocities of the surface
states of SmB$_{6}$ 
measured in experiment.
Both quantum oscillation\cite{Li13} and ARPES
studies\cite{Xu13,Jiang13,Neupane13}, show the surface quasiparticles
are light. ARPES measurements indicate surface quasiparticles
 with Fermi velocities ranging from 220$
{\rm meV \AA}$\cite{Jiang13} to 300
${\rm meV\AA}$̊\cite{Neupane13}. Velocities obtained from dHvA
measurements are two orders of magnitude larger than these values\cite{Li13}. 
By contrast, current
theories\cite{dzerotki2,Alexandrov13,Lu13,BitanRoy} predict heavy Dirac
quasiparticles with velocities vs $\sim 30-50{\rm meV\AA}$. 
Alexandrov, Coleman and Erten \cite{Erten15} have recently suggested 
a link between these discrepancies and a possible break-down of the
Kondo effect on the surface. 
They argue that the  reduced co-ordination
numbers of magnetic ions on the surface, the Kondo temperature is 
suppressed, leading to surface {\sl Kondo breakdown}. The re-emergence
of unscreened local moments at the surface 
is likely to favor the development of surface magnetism.
Alexandrov et al. find that the release of d-electrons from Kondo
singlets at the surface does not destroy the surface states, but
instead
has the effect of doping the surface Dirac
cones and driving the Dirac point into the continuum;
surface Kondo breakdown also 
increases the conduction band character 
of the topological surface  states,  leading to an approximately 
ten fold increase in the quasiparticle velocities, a result that may
account for the high surface
velocities seen in ARPES measurements on SmB$_{6}$. 

The history of \SMB{} studies has invariably shown that new physics 
accompanies higher quality samples. 
Currently, \SMB{} crystals are usually grown
using either the aluminum flux method or the floating zone method. The
majority of the reviewed experiments have been performed on 
flux-grown-samples, while high quality floating-zone-grown samples
also showed resistance saturation as reported by Hatnean et
al. ~\cite{Hatnean13} and numerous older reports. However, the
materials aspect of \SMB{} is far from trivial. 
Phelan et al. ~\cite{Phelan14} recently report that 
some floating-zone-grown
\SMB{} crystals 
remain insulating at the lowest temperature measured, while
additional non-magnetic carbon doping restores the surface metallic
state. 
While the exact role of carbon doping is unclear at this stage,
these result clearly highlight the complexity of surface chemistry in
\SMB{} and the vast number of possibilities of controlling the
topological and non-topological properties of \SMB{}.

\section*{DISCLOSURE STATEMENT}
The authors are not aware of any affiliations, memberships, funding, or financial holdings that might be perceived as 
affecting the objectivity of this review.

\section*{ACKNOWLEDGMENTS}
This work was supported by the Ohio Board of Regents Research
Incentive Program grant OBR-RIP-220573 nd MPI-PKS (M.D.), DARPA and
Simons Foundation (V.G.), and DOE grant DE-FG02-99ER45790 (P.C.) and
the  U.S. National Science Foundation I2CAM International Materials
Institute Award, Grant DMR-1411344 (P.C.). 
We gratefully acknowledge valuable discussions with Jim Allen, Onur Erten, 
Gilbert Lonzarich, Suchitra Sebastian, Kai Sun and Zachary Fisk.

\bibliographystyle{ar-style4}

\end{document}